\documentclass[12pt]{article}

\usepackage{draft} 
\usepackage{hyperref}
\usepackage{graphicx,color}
\usepackage{cite}
\usepackage{mciteplus}
\usepackage{skak}
\usepackage{empheq}
\usepackage{tikz}
\usepackage{booktabs}
\usepackage{siunitx}
\usepackage{bbm}
\usepackage{makecell}
\usepackage{float}
\usepackage{dsfont}

\usepackage{bm}
\usepackage{braket}
\usepackage{subcaption}

\DeclareFontFamily{OT1}{pzc}{}
\DeclareFontShape{OT1}{pzc}{m}{it}{<-> s * [1.10] pzcmi7t}{}
\DeclareMathAlphabet{\mathpzc}{OT1}{pzc}{m}{it}

\def\be#1\ee{\begin{align}#1\end{align}}

\makeatletter
\newcommand{\bdryno}{\mathpalette\bdry@no\relax}
\newcommand{\bdry@no}[2]{%
  \mspace{1mu}%
  \vbox{%
    \hbox{$\m@th#1\scriptstyle{\ast}$}
    \nointerlineskip
    \kern.25ex
    \hbox{$\m@th#1\scriptstyle{\ast}$}
    \kern-.06ex
  }%
  \mspace{1mu}%
}
\makeatother

\usetikzlibrary{snakes}
\usetikzlibrary{arrows}
\usetikzlibrary{decorations.pathmorphing}
\usetikzlibrary{ decorations.markings}
\tikzset{snake it/.style={decorate, decoration=snake}}
\usetikzlibrary{shapes.misc}
\tikzset{cross/.style={cross out, draw=black, minimum size=2*(#1-\pgflinewidth), inner sep=0pt, outer sep=0pt},
cross/.default={1pt}}
\usetikzlibrary{shapes.geometric}

\definecolor{bleudefrance}{rgb}{0.19, 0.55, 0.91}
\definecolor{candyapplered}{rgb}{1.0, 0.03, 0.0}
\definecolor{myblue}{rgb}{0.0,0.635,1}

\newcommand{\ups}{\Upsilon_{1}}

\newcommand{\zbar}{\overline{z}}
\newcommand{\cF}{\mathcal{F}}

\newcommand{\Chat}{\widehat{C}}
\newcommand{\Vhat}{\widehat{V}}
\newcommand{\Phat}{\widehat{P}}
\newcommand{\hhat}{\widehat{h}}

\tikzset{
  pics/cylnnrb0/.style n args={2}{
    code = { %
        \filldraw[color=red,fill=black!30, thick] circle (0.35);
        \filldraw[color=blue, fill=white, thick]  circle (0.17);
        \node[color=black] at (0,0.51) {\scriptsize ${#1}$};
        \node[color=black] at (0,0) {\scriptsize ${#2}$};

    }
  }
}

\tikzset{
  pics/cylnnrb/.style n args={2}{
    code = { %
        \filldraw[color=red, densely dashed,fill=black!30, thick] circle (0.35);
        \filldraw[color=blue, densely dashed, fill=white, thick]  circle (0.17);
        \node[color=black] at (0,0.51) {\scriptsize ${#1}$};
        \node[color=black] at (0,0) {\scriptsize ${#2}$};

    }
  }
}
\tikzset{
  pics/cylnnrr0/.style n args={2}{
    code = { %
        \filldraw[color=red,fill=black!30, thick] circle (0.35);
        \filldraw[color=red, fill=white, thick]  circle (0.17);
        \node[color=black] at (0,0.51) {\scriptsize ${#1}$};
        \node[color=black] at (0,0) {\scriptsize ${#2}$};

    }
  }
}
\tikzset{
  pics/cylnnrr/.style n args={2}{
    code = { %
        \filldraw[color=red, densely dashed,fill=black!30, thick] circle (0.35);
        \filldraw[color=red, densely dashed, fill=white, thick]  circle (0.17);
        \node[color=black] at (0,0.51) {\scriptsize ${#1}$};
        \node[color=black] at (0,0) {\scriptsize ${#2}$};

    }
  }
}
\tikzset{
  pics/cylnnrr1/.style n args={2}{
    code = { %
        \filldraw[color=red,fill=black!30, thick] circle (0.35);
        \filldraw[color=red, fill=white, thick]  circle (0.17);
        \node[color=black] at (0,0.51) {\scriptsize ${#1}$};
        \node[color=black] at (0,0) {\scriptsize ${#2}$};
        \node[cross=3pt, very thick] at (-0.25,0) {};
    }
  }
}
\tikzset{
  pics/cylnnbb0/.style n args={2}{
    code = { %
        \filldraw[color=blue,fill=black!30, thick] circle (0.35);
        \filldraw[color=blue, fill=white, thick]  circle (0.17);
        \node[color=black] at (0,0.51) {\scriptsize ${#1}$};
        \node[color=black] at (0,0) {\scriptsize ${#2}$};

    }
  }
}
\tikzset{
  pics/cylnnbb/.style n args={2}{
    code = { %
        \filldraw[color=blue, densely dashed,fill=black!30, thick] circle (0.35);
        \filldraw[color=blue, densely dashed, fill=white, thick]  circle (0.17);
        \node[color=black] at (0,0.51) {\scriptsize ${#1}$};
        \node[color=black] at (0,0) {\scriptsize ${#2}$};

    }
  }
}
\tikzset{
  pics/cylnnbb1/.style n args={2}{
    code = { %
        \filldraw[color=blue,fill=black!30, thick] circle (0.35);
        \filldraw[color=blue, fill=white, thick]  circle (0.17);
        \node[color=black] at (0,0.51) {\scriptsize ${#1}$};
        \node[color=black] at (0,0) {\scriptsize ${#2}$};
        \node[cross=3pt, very thick] at (-0.25,0) {};
    }
  }
}

\tikzset{
  pics/disk1/.style n args={1}{
    code = { %
        \filldraw[color=black, fill=black!30, thick] circle (0.3);
        \node[color=black] at (0,0.5) {\scriptsize ${#1}$};
        \draw node[cross=3pt, very thick] {};
    }
  }
}   
\tikzset{
  pics/disk1r/.style n args={1}{
    code = { %
        \filldraw[color=red, fill=black!30, thick] circle (0.3);
        \node[color=black] at (0,0.5) {\scriptsize ${#1}$};
        \draw node[cross=3pt, very thick] {};
    }
  }
}
\tikzset{
  pics/disk2r/.style n args={1}{
    code = { %
        \filldraw[color=red, fill=black!30, thick] circle (0.3);
        \node[color=black] at (0,0.5) {\scriptsize ${#1}$};
        \node[cross=3pt, very thick] at (0.125,0) {};
        \node[cross=3pt, very thick] at (-0.125,0) {};
    }
  }
}
\tikzset{
  pics/disker/.style n args={1}{
    code = { %
        \filldraw[color=red, fill=black!30, thick] circle (0.3);
        \node[color=black] at (0,0.5) {\scriptsize ${#1}$};
    }
  }
}
\tikzset{
  pics/disk1b/.style n args={1}{
    code = { %
        \filldraw[color=blue, fill=black!30, thick] circle (0.3);
        \node[color=black] at (0,0.5) {\scriptsize ${#1}$};
        \draw node[cross=3pt, very thick] {};
    }
  }
}   
\tikzset{
  pics/disk2b/.style n args={1}{
    code = { %
        \filldraw[color=blue, fill=black!30, thick] circle (0.3);
        \node[color=black] at (0,0.5) {\scriptsize ${#1}$};
        \node[cross=3pt, very thick] at (0.125,0) {};
        \node[cross=3pt, very thick] at (-0.125,0) {};
    }
  }
} 
\tikzset{
  pics/diskeb/.style n args={1}{
    code = { %
        \filldraw[color=blue, fill=black!30, thick] circle (0.3);
        \node[color=black] at (0,0.5) {\scriptsize ${#1}$};
    }
  }
}

\begin{document}

\unitlength = .8mm

\begin{titlepage}

\begin{center}

\hfill \\
\hfill \\
\vskip 1cm


\title{A Two-Dimensional String Cosmology}

\author{Victor A. Rodriguez}

\address{
Joseph Henry Laboratories, Princeton University, \\ Princeton, NJ 08544, USA 
}

\email{vrodriguez@princeton.edu}

\end{center}

\abstract{
We study two-dimensional string theory on a time-dependent background, whose worldsheet description consists of Liouville theory at central charge $c=1$ and Liouville theory at central charge $c=25$, together with the conformal ghosts. We compute the tree-level three-point and four-point components of the cosmological wavefunction in string perturbation theory. 
The latter is evaluated numerically by decomposing the Liouville four-point correlation functions into Virasoro conformal blocks and three-point function coefficients and integrating over the moduli space of the four-punctured sphere string diagram.
This computation numerically confirms a surprisingly simple conjectural result for the four-point wavefunction component whose physical interpretation remains to be clarified. 
}

\vfill

\end{titlepage}

\eject

\begingroup
\hypersetup{linkcolor=black}

\tableofcontents

\endgroup

\pagebreak


\section{Introduction}

A difficult problem in string theory is to construct time-dependent backgrounds, or cosmological spacetimes, that resemble the evolution of our universe. The hope is that such string-theoretic models would help provide a microscopic understanding of the origin of our universe and the dynamics of particles at early cosmological times.

Two-dimensional string theories provide rich laboratories for investigating fundamental aspects of string theory in the simplest possible context (for reviews, see \cite{Klebanov:1991qa,Ginsparg:1993is,Jevicki:1993qn,Polchinski:1994mb,Martinec:2004td,Nakayama:2004vk}).
They exhibit interesting phenomena such as holographic duality with matrix quantum mechanics \cite{Brezin:1989ss,Gross:1990ay,Ginsparg:1990as,Douglas:2003up,Takayanagi:2003sm}, non-perturbative effects mediated by D-instantons \cite{Balthazar:2019rnh,Balthazar:2019ypi,Sen:2019qqg,Sen:2020oqr,Sen:2020ruy,Sen:2020eck,Sen:2021qdk,DeWolfe:2003qf,Balthazar:2022apu,Chakravarty:2022cgj,Sen:2022clw,Eniceicu:2022xvk}, and open string time-dependent dynamics of rolling tachyons\footnote{Open string rolling tachyons are time-dependent phenomena that are localized to the boundary of the string worldsheet. Instead, in this paper we are interested in time-dependence that affects the full two-dimensional worldsheet of the closed strings, and that is present in the entire target spacetime of the theory.} \cite{Sen:2004nf,McGreevy:2003kb,McGreevy:2003ep,Klebanov:2003km}. The purpose of the present work is to investigate a simple model of closed strings interacting in a time-dependent two-dimensional spacetime.

More precisely, we investigate a closed string time-dependent background that admits an exact worldsheet conformal field theory (CFT) description. Concretely, the worldsheet CFT of the two-dimensional string cosmology model is the following
\begin{equation}
\parbox{.2\textwidth}{\centering $c=1$ \\ Liouville CFT} \oplus \parbox{.2\textwidth}{\centering $c=25$ \\ Liouville CFT} \oplus \parbox{.2\textwidth}{\centering $b,c$ ghosts,}
\label{eq:wscft}
\end{equation}
where the Liouville sector at central charge $c=1$, suitably continued to Lorentzian signature as explained in Section \ref{sec:wscosmo}, is interpreted as time, and where the Liouville sector at central charge $c=25$ is interpreted as space as in the more familiar (time-independent) two-dimensional bosonic or type 0B/0A string theories. 

While in this paper we will adopt the bootstrap approach to describe the worldsheet theory (\ref{eq:wscft}) as an abstract CFT in terms of its operator spectrum and three-point function coefficients, it is useful to consider the worldsheet action of Liouville theory in order to understand the spacetime interpretation of the string theory. The action for Liouville theory is given by
\ie
S_{\text{Liouv.}}[\phi] = \frac{1}{4\pi} \int d^2\sigma	\sqrt{g} \left( g^{mn}\partial_m\phi\partial_n\phi + QR\phi + 4\pi\mu e^{2b\phi} \right),
\label{eq:LiouvAction}
\fe
where $\phi$ is the Liouville field, $R$ is the Ricci scalar of the two-dimensional worldsheet of the string with metric $g$, and $\mu$ is a constant\footnote{A rescaling of the constant $\mu$ corresponds to a rescaling of the string coupling $g_s$.}. The central charge of the theory is related to the background charge $Q$ of the linear dilaton term by $c=1+6Q^2$, which in turn is related to the parameter $b$ that controls the Liouville exponential potential by $Q=b+b^{-1}$.

For the $c=25$ Liouville sector of the worldsheet CFT (\ref{eq:wscft}) we have that $b=1$, $Q=2$. The string coupling $g_s$ is controlled by the linear dilaton term that multiplies the Ricci scalar $R$, and thus goes as $g_s\sim e^{2\phi}$. Since the field $\phi$ is interpreted as the spatial coordinate of the two-dimensional spacetime, the strength of the string coupling varies exponentially in space. The potentially uncontrolled region of strong string coupling as $\phi\to\infty$, however, is shielded by the exponential ``Liouville wall" $4\pi\mu e^{2\phi}$ that reflects closed strings back to the asymptotic region of weak string coupling at $\phi\to -\infty$.

On the other hand, for the $c=1$ Liouville sector of (\ref{eq:wscft}) the relevant value for the Liouville parameter is $b=i$. 
To distinguish this $c=1$ sector, let us denote its Liouville field by $\varphi$. 
The first immediate consequence is that $Q=0$ in this case and therefore, since this $c=1$ sector will be interpreted as the time coordinate of the spacetime, the string coupling $g_s$ is independent of time. 
Furthermore, in order to interpret the worldsheet field $\varphi$ as a Lorentzian time coordinate in target spacetime we perform a Wick rotation in field space such that $\varphi = i \chi^0$, and the action becomes
\ie
S_{c=1\text{ Liouv.}}[\chi^0] = \frac{1}{4\pi} \int d^2\sigma	\sqrt{g} \left( -g^{mn}\partial_m\chi^0\partial_n\chi^0 + 4\pi\mu e^{-2\chi^0} \right).
\label{eq:timeLiouv}
\fe
The last term of (\ref{eq:timeLiouv}) represents a time-dependent potential in spacetime that grows exponentially large in the infinite past $\chi^0 \to -\infty$. In the far future, the potential vanishes and we recover the familiar two-dimensional bosonic string vacuum, whose $c=1$ sector is a time-like free boson.

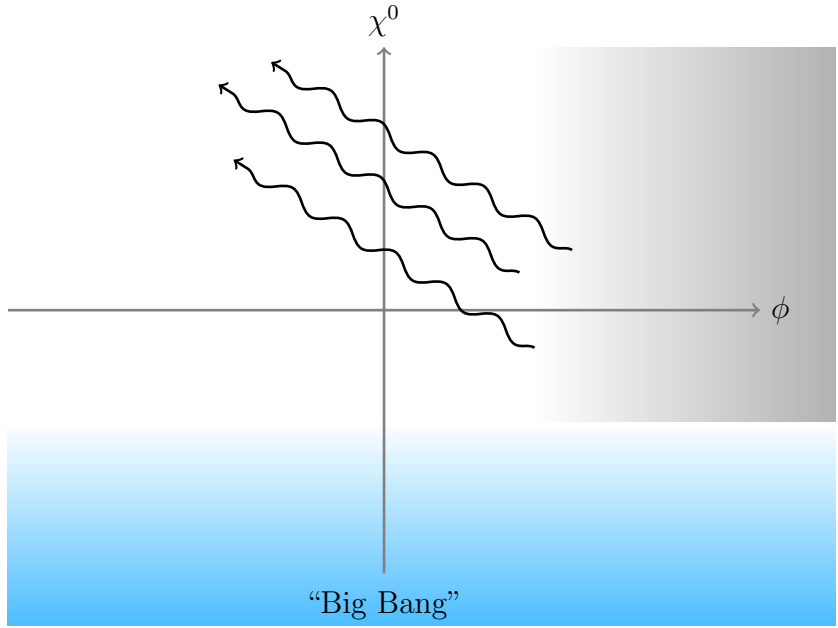
\begin{figure}[h!]
\centering
\begin{tikzpicture}
\shade[left color = black!0,right color = black!30] (2,-3.5) rectangle (6,3.5);
\shade[top color = myblue!0, bottom color = myblue!70] (-5,-4.2) rectangle (6,-1.5);
\draw[->,color=gray, line width=1pt] (-5,0) -- (5,0);
\node[right] at (5,0) {$\phi$};
\draw[->,color=gray, line width=1pt] (0,-3.5) -- (0,3.5);
\node[above] at (0,3.5) {$\chi^0$};
\draw[->, color=black, line width=1pt, decorate, decoration={snake, segment length=8mm, amplitude=1mm}] (2,-0.5) -- (-2,2);
\draw[->, color=black, line width=1pt, decorate, decoration={snake, segment length=8mm, amplitude=1mm}] (1.8,0.5) -- (-2.2,3);
\draw[->, color=black, line width=1pt, decorate, decoration={snake, segment length=8mm, amplitude=1mm}] (2.5,0.8) -- (-1.5,3.3);
\node[above] at (0,-4.25) {``Big Bang"};
\end{tikzpicture}
\caption{Spacetime interpretation of the two-dimensional string cosmology background. The region shaded in gray represents the space-like Liouville potential wall, whereas the region shaded in blue represents the time-dependent Liouville potential $e^{-2\chi^0}$. Worldsheet string diagrams in string perturbation theory compute the overlap of an initial state in the infinite past, denoted by ``Big Bang", and an out-state in the Hilbert space of perturbative string states, i.e. a particular component of the cosmological wavefunction.}
\label{fig:spacetime}
\end{figure}

The spacetime interpretation of the two-dimensional string theory background described by the worldsheet CFT (\ref{eq:wscft}) is pictured in Figure \ref{fig:spacetime}. 
The spatial dimension parametrized by the $c=25$ Liouville field $\phi$ runs horizontally and the time direction parametrized by the $c=1$ Liouville field $\chi^0$ runs vertically. The region shaded with a gradient in grey represents the space-like Liouville wall that shields the strong string coupling region, and the region shaded with a gradient in blue represents the time-dependent exponential potential in (\ref{eq:timeLiouv}). 

Worldsheet string perturbation theory in this case does not compute an S-matrix element of in- and out-states in the Hilbert space of perturbative string states. Instead, we propose that it computes the overlap of an initial state of the two-dimensional spacetime in the infinite past, labeled as ``Big Bang" in Figure \ref{fig:spacetime}, with an out-state in the Hilbert space of string states --- a particular component of the cosmological wavefunction in the basis of perturbative string states in the far future.

In this paper, we compute the simplest nontrivial cosmological wavefunction components in the time-dependent background (\ref{eq:wscft}), the three- and four-point wavefunction components at tree-level in string perturbation theory. We will not, however, make use of the semiclassical Liouville action (\ref{eq:LiouvAction}). The two formal manipulations we performed in order to arrive at (\ref{eq:timeLiouv}), namely the continuation of the Liouville parameter $b\to i$ and the rotation in field space to Lorentzian time in target space $\varphi\to i \chi^0$, will be described in terms of the abstract CFT structure of $c=1$ Liouville theory.

Although the string-theoretic calculation of cosmological wavefunction components is technically complicated, particularly the four-point string diagram discussed in Section \ref{sec:4ptwavef}, we find strikingly simple expressions for the three-point and four-point cosmological wavefunction components in terms of the outgoing energies of the asymptotic closed strings in the infinite future. 
The tree-level cosmological wavefunction components (\ref{eq:3ptWavef}) and (\ref{eq:myfit}) are the main result of this paper, but both their simplicity and a more fundamental understanding of their physical interpretation remain to be clarified. We will leave this to future work.

The rest of the paper is organized as follows. In Section \ref{sec:wstheory}, we review the bootstrap results of Liouville CFT at $c=1$ and at $c=25$, and then describe the asymptotic string states (in the far future) of the worldsheet string theory (\ref{eq:wscft}). 
In Section \ref{sec:cosmowavef}, we present the computation of the three- and four-point cosmological wavefunction components of the two-dimensional string cosmology. We conclude in Section \ref{sec:discussion} with a discussion of our results and future directions.

\section{Worldsheet description of two-dimensional string cosmology}
\label{sec:wstheory}

In Sections \ref{sec:c25Liouv} and \ref{sec:c1Liouv}, we describe Liouville theory at central charges $c=25$ and $c=1$ as abstract conformal field theories, in terms of their operator spectrum and their respective three-point function coefficients subject to the consistency conditions of crossing symmetry of the four-point function on the sphere and the modular covariance of the one-point function on the torus. 
In Section \ref{sec:wscosmo}, we combine these ingredients to describe the time-dependent string theory background of Figure \ref{fig:spacetime}.

\subsection{Liouville CFT at $c=25$}
\label{sec:c25Liouv}

The complete operator spectrum of Liouville theory at central charge $c=25$ consists of a continuum of scalar Virasoro primaries $V_P$, labeled by the ``Liouville momentum" $P\in \bR_{\geq 0}$ and with conformal weights $h=\widetilde{h}=1+P^2$, together with its Virasoro descendants. In this paper, we will follow the normalization convention of \cite{Balthazar:2017mxh} in which the vertex operators $V_P$ are delta-function normalized\footnote{In this convention, the wavefunctional of the vertex operator $V_P$ in the asymptotic region $\phi\to -\infty$ of Liouville target space is identified with the free field expression
\ie
V_P \sim S(P)^{-\frac{1}{2}} e^{(2+2iP)\phi} + S(P)^{\frac{1}{2}} e^{(2-2iP)\phi},
\fe
where the reflection phase $S(P) = -(\Gamma(2iP)/\Gamma(-2iP))^2$. With this convention, there are no additional ``leg-pole" factors when comparing perturbative string amplitudes in $c=1$ string theory and amplitudes in the $c=1$ matrix quantum mechanics \cite{Balthazar:2017mxh}.
},
\ie
\langle V_P(z,\zbar) V_{P'}(0) \rangle = \pi \frac{\delta(P-P')}{|z|^{4h}}.
\label{eq:LiouvNorm}
\fe

The three-point function coefficients, defined by
\ie
\langle V_{P_1}(z_1,\zbar_1) V_{P_2}(z_2,\zbar_2) V_{P_3}(z_3,\zbar_3) \rangle = \frac{C(P_1,P_2,P_3)}{|z_{12}|^{2(h_1+h_2-h_3)}|z_{13}|^{2(h_1+h_3-h_2)}|z_{23}|^{2(h_2+h_3-h_1)}},
\label{eq:3ptc25}
\fe
(where $h_i\equiv 1+P_i^2$)
were bootstrapped in \cite{Dorn:1994xn, Zamolodchikov:1995aa} and are given by the well-known DOZZ formula which takes the following form for the particular value of the central charge $c=25$,
\ie
C(P_1,P_2,P_3) = \frac{1}{\ups(1+i(P_1+P_2+P_3))} \left[ \frac{2P_1\ups(1+2iP_1)}{\ups(1+i(P_2+P_3-P_1))}\times(\text{2 permutations}) \right],
\label{eq:DOZZ1}
\fe
where the function $\ups(x)$ is a special case of Barnes double Gamma function, and is defined as
\ie
\ups(x) = \frac{1}{\Gamma_1(x)\Gamma_1(2-x)},
\label{eq:defUps1}
\fe
and where $\Gamma_1(x)$ is related to the Barnes G-function $G(x)$ by $\Gamma_1(x) = (2\pi)^{(x-1)/2}(G(x))^{-1}$ \cite{Balthazar:2018qdv}. 
$\Upsilon_1(x)$ is an entire function with zeros at $x=n$ with multiplicity $(n-1)$ for $n\in\bZ_{\geq 2}$ and at $x=-m$ with multiplicity $(m+1)$ for $m\in\bZ_{\geq 0}$. Note that the three-point function coefficient (\ref{eq:DOZZ1}) is real-valued for real Liouville momenta $P_i$.

The four-point function of $c=25$ Liouville vertex operators on the sphere admits a decomposition in terms of the three-point function coefficients (\ref{eq:DOZZ1}) and $c=25$ Virasoro conformal blocks as
\ie
\langle & V_{P_4}(\infty) V_{P_3}(1) V_{P_2}(z,\zbar) V_{P_1}(0) \rangle_{c=25\text{ Liouv.}} \vphantom{\int}\\
&= \int_0^\infty \frac{dP}{\pi} C(P_1,P_2,P)C(P_3,P_4,P) \cF_{c=25}(h_4,h_3,h_2,h_1;1+P^2|z)\cF_{c=25}(h_4,h_3,h_2,h_1;1+P^2|\zbar),
\label{eq:4ptLiouvc25}
\fe
where $\cF_{c}(h_4,h_3,h_2,h_1;h|z)$ is the sphere four-point holomorphic Virasoro conformal block at central charge $c$ with external weights $h_i$ for $i=1,\ldots,4$, intermediate weight $h$, evaluated at the cross-ratio $z$.
Similarly, the one-point function on the torus with modulus $\tau$ can be decomposed into Virasoro conformal blocks as
\ie
\left\langle V_{P_{\rm ext}}(0) \vphantom{\Vhat}\right\rangle^{T^2(\tau)}_{c=25 \text{ Liouv.}} = \int_0^\infty \frac{dP}{\pi} C(P_{\rm ext},P,P) \cF_{c=25}(h_{\rm ext};1+P^2;q) \cF_{c=25}(h_{\rm ext};1+P^2;\overline{q}),
\label{eq:torusCBdecompc25}
\fe
where $\cF_{c}(h_{\rm ext};h_{\rm int};q)$ is the torus one-point Virasoro conformal block at central charge $c$ with external weight $h_{\rm ext}$, internal weight $h_{\rm int}$, and $q=e^{2\pi i \tau}$. Our conventions for the sphere and torus Virasoro conformal blocks are given in Appendix \ref{sec:recursions}. In Appendix \ref{sec:consistency}, we verify numerically the crossing symmetry of sphere four-point function (\ref{eq:4ptLiouvc25}) and the modular covariance of torus one-point function (\ref{eq:torusCBdecompc25}).

\subsection{Liouville CFT at $c=1$}
\label{sec:c1Liouv}

The CFT data of Liouville theory at $c=1$ is similar to that of $c=25$ presented in the previous section. Naively, Liouville theory at central charge $c\leq 1$ is obtained by the continuation of the Liouville parameter $b$ to imaginary values $b=i\beta$, with $\beta\in\bR$. 
The analytic continuation of $b$ to complex values, however, fails precisely along the imaginary axis. For instance, the upsilon function $\Upsilon_b(x)$ in terms of which the structure constants are written diverges in this limit \cite{Ribault:2015sxa}. Consequently, the bootstrap of the structure constants along this branch of $b=i\beta$ has to be reanalyzed. This has been done in \cite{Schomerus:2003vv,Kostov:2005kk,Zamolodchikov:2005fy,Ribault:2015sxa}\footnote{See \cite{Harlow:2011ny,McElgin:2007ak,Giribet:2011zx} as well.}, and we will present the result for the particular case of $c=1$ ($\beta=1$).

The operator spectrum of Liouville CFT at $c=1$ again consists of a continuum of scalar Virasoro primaries $\Vhat_P$, labeled by a momentum $P\in\bR$ and now with conformal weights $\hhat = \widetilde{\hhat} = P^2$, together with its Virasoro descendants.\footnote{In what follows we will denote by a hat quantities that refer to Liouville CFT at $c=1$ in order to distinguish those of Liouville CFT at $c=25$.} 
We will also assume the vertex operators $\Vhat_P$ are delta-function normalized,\footnote{In this case, the Liouville reflection phase in the limit $\beta\to 1$ becomes precisely unity, $\widehat{S}(P)=1$.} 
\ie
\langle \Vhat_P(z,\zbar) \Vhat_{P'}(0) \rangle = \pi \frac{\delta(P-P')}{|z|^{4\hhat}}.
\fe
In turn, the three-point function coefficients that appear in
\ie
\langle \Vhat_{P_1}(z_1,\zbar_1) \Vhat_{P_2}(z_2,\zbar_2) \Vhat_{P_3}(z_3,\zbar_3) \rangle = \frac{\Chat(P_1,P_2,P_3)}{|z_{12}|^{2(\hhat_1+\hhat_2-\hhat_3)}|z_{13}|^{2(\hhat_1+\hhat_3-\hhat_2)}|z_{23}|^{2(\hhat_2+\hhat_3-\hhat_1)}},
\fe
are given by
\ie
\Chat(P_1,P_2,P_3) = \ups(1+P_1+P_2+P_3) \left[ \frac{\ups(1+P_2+P_3-P_1)}{\ups(1+2P_1)}\times(\text{2 permutations}) \right].
\label{eq:DOZZ2}
\fe
Note that (\ref{eq:DOZZ2}) is real-valued for real or purely imaginary Liouville momenta $P_i\in\bR,i\bR$, and that (\ref{eq:DOZZ2}) is an even function of any of the momenta $P_i$.


The four-point function of Liouville vertex operators in $c=1$ Liouville CFT admits the following Virasoro conformal block decomposition,
\ie
\langle & \Vhat_{P_4}(\infty) \Vhat_{P_3}(1) \Vhat_{P_2}(z,\zbar) \Vhat_{P_1}(0) \rangle_{c=1\text{ Liouv.}} \vphantom{\int}\\
&= \int_\cC \frac{d\Phat}{2\pi} \Chat(P_1,P_2,\Phat)\Chat(P_3,P_4,\Phat) \cF_{c=1}(\hhat_4,\hhat_3,\hhat_2,\hhat_1;\Phat^2|z)\cF_{c=1}(\hhat_4,\hhat_3,\hhat_2,\hhat_1;\Phat^2|\zbar),
\label{eq:4ptLiouvc1}
\fe
where the contour of integration $\cC$ over the intermediate state with Liouville momentum $\Phat$ is specified as follows.
In contrast with the DOZZ structure constant (\ref{eq:DOZZ1}), $\Chat(P_1,P_2,\Phat)$ viewed as a function over the complex $\Phat$-plane only has poles (not necessarily simple) at $\Phat = {k \over 2}$ for $k\in\bZ_{\neq 0}$.
In addition, the $c=1$ Virasoro conformal block $\cF_{c=1}(\hhat_4,\hhat_3,\hhat_2,\hhat_1;\Phat^2|z)$ has poles at $\Phat = {n \over 2}$ for $n\in\bZ$. 
The contour prescription $\cC$ in (\ref{eq:DOZZ2}) is chosen to run parallel to the real axis and shifted vertically by a small $\epsilon > 0$ amount to avoid the poles as shown in Figure \ref{fig:contourPhat} \cite{Ribault:2015sxa}. 
In Appendix \ref{sec:consistency}, we verify numerically that the four-point function (\ref{eq:4ptLiouvc1}) satisfies crossing symmetry with this contour prescription.

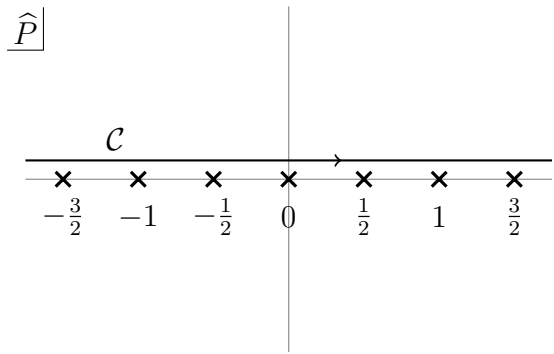
\begin{figure}[h]
\centering
\begin{tikzpicture}
[decoration={markings,mark=at position 0.6 with {\arrow{>}}}] 
\draw[color=gray] (-3.5,0) -- (3.5,0);
\draw[color=gray] (0,-2.3) -- (0,2.3);
\draw (0,0) node[cross=4pt, very thick] {};
\node at (0,-0.5) {$0$};
\draw (1,0) node[cross=4pt, very thick] {};
\node at (1,-0.5) {$\frac{1}{2}$};
\draw (2,0) node[cross=4pt, very thick] {};
\node at (2,-0.5) {$1$};
\draw (3,0) node[cross=4pt, very thick] {};
\node at (3,-0.5) {$\frac{3}{2}$};
\draw (-1,0) node[cross=4pt, very thick] {};
\node at (-1,-0.5) {$-\frac{1}{2}$};
\draw (-2,0) node[cross=4pt, very thick] {};
\node at (-2,-0.5) {$-1$};
\draw (-3,0) node[cross=4pt, very thick] {};
\node at (-3,-0.5) {$-\frac{3}{2}$};
\draw[thick,postaction={decorate}] (-3.5,0.25) -- (3.5,0.25);
\draw (-2.3,0.25) node[above] {$\cC$};
\node(n)[inner sep=2pt] at (-3.5,2) {$\Phat$};
\draw[line cap=round](n.south west)--(n.south east)--(n.north east);
\end{tikzpicture}
\caption{Contour of integration $\cC$ over the intermediate states in the Virasoro conformal block decomposition of the four-point function (\ref{eq:4ptLiouvc1}) in Liouville CFT at $c=1$. Poles of the $\Phat$-integrand are marked with crosses. The contour $\cC$ runs parallel to the real axis and shifted vertically by a small $\epsilon>0$ amount in the imaginary direction in order to avoid the poles. The integration over intermediate states in the Virasoro conformal block decomposition of the torus one-point function (\ref{eq:torusCBdecompc1}) has the same pole structure, without counting multiplicities, and hence we use the same contour prescription.}
\label{fig:contourPhat}
\end{figure}


Likewise, the one-point function on a torus with modulus $\tau$ in $c=1$ Liouville CFT admits the conformal block decomposition,
\ie
\left\langle \Vhat_{P_{\rm ext}}(0) \vphantom{\Vhat}\right\rangle^{T^2(\tau)}_{c=1 \text{ Liouv.}} = \int_\cC \frac{d\Phat}{2\pi} \Chat(P_{\rm ext},\Phat,\Phat) \cF_{c=1}(\hhat_{\rm ext};\Phat^2;q) \cF_{c=1}(\hhat_{\rm ext};\Phat^2;\overline{q}).
\label{eq:torusCBdecompc1}
\fe
In particular, the poles of the integrand on the RHS of (\ref{eq:torusCBdecompc1}) coming from $\Chat(P_{\rm ext},\Phat,\Phat)$ again occur at $\Phat = {m\over 2}$ for $m\in\bZ_{\neq 0}$, whereas those coming from the torus one-point conformal block $\cF_{c=1}(\hhat_{\rm ext};\Phat^2;q)$ occur at $\Phat = {n \over 2}$ for $n\in\bZ$. Therefore, the integrand on the RHS of (\ref{eq:torusCBdecompc1}) has the same pole structure as shown in Figure \ref{fig:contourPhat} (not counting multiplicities of the poles), and the same contour prescription $\cC$ is used in (\ref{eq:torusCBdecompc1}) as well. 
In Appendix \ref{sec:consistency}, we verify numerically that the torus one-point function (\ref{eq:torusCBdecompc1}) with the contour prescription shown in Figure \ref{fig:contourPhat} is modular covariant.\footnote{In two-dimensional CFT, the sum/integration over intermediate operators in an operator product expansion, or conformal block, decomposition of a correlation function such as (\ref{eq:4ptLiouvc1}) or (\ref{eq:torusCBdecompc1}) is performed over operators strictly in the spectrum of the CFT, which we took to be $\Vhat_{\Phat}$ with $\Phat\in\bR$. Here, we relax this CFT axiom and shift the contour of integration as in Figure \ref{fig:contourPhat} in order to have a well-defined correlation function, while maintaining crossing symmetry of the sphere four-point function and modular covariance of the torus one-point function.}

\subsection{Two-dimensional string cosmology}
\label{sec:wscosmo}

The full worldsheet CFT description of the two-dimensional string cosmology background consists of the Liouville CFTs at $c=1$ and $c=25$, together with the $b,c$ conformal ghosts.

The Liouville CFT at central charge $c=1$ described in Section \ref{sec:c1Liouv} describes a theory in Euclidean signature. In order to have a description of the time-dependent background in Lorentzian signature shown in Figure \ref{fig:spacetime}, we will make one further modification to the worldsheet CFT. We will analytically continue the $c=1$ Euclidean Liouville momenta $P$ of \emph{external} on-shell string states to imaginary values,
\ie
\Vhat_{P\to -i \omega},
\label{eq:Lorcontinuation}
\fe
such that $\omega\in \bR_{\geq 0}$ has the interpretation of a Lorentzian energy of the string asymptotic state. 
With this continuation of the $c=1$ Liouville CFT sector, on-shell asymptotic states in the two-dimensional string cosmology background corresponding to BRST cohomology classes of the full worldsheet CFT are represented by the following vertex operators,
\ie
\cV_\omega \,=\, g_s\, \Vhat_{-i\omega} \, V_{P=\omega}.
\label{eq:vertexop}
\fe
As in other (time-independent) two-dimensional string theories, the on-shell condition that $\cV_\omega$ is a weight $(1,1)$ vertex operator leads to a dispersion relation of a massless scalar in $1+1$ dimensions: $-\omega^2 + P^2 = 0$ where the spatial $c=25$ Liouville momentum takes the value $P=\omega$.


The continuation (\ref{eq:Lorcontinuation}) in $c=1$ Liouville CFT can be justified at the level of the sphere four-point function as follows. 
Consider again the Virasoro conformal block decomposition in (\ref{eq:4ptLiouvc1}). We want to continue the external $c=1$ Liouville momenta $P_i\to -i\omega_i$ for $i=1,\ldots,4$, while maintaining analyticity of the CFT four-point function, deforming the contour of integration $\cC$ over the intermediate state with momentum $\Phat$ accordingly, if necessary. 
The key property of the $c=1$ structure constant $\Chat(P_1,P_2,\Phat)$ is that the positions of its poles in the complex $\Phat$-plane are \emph{independent} of the values of the external momenta $P_1$ and $P_2$, and hence the analytic continuation (\ref{eq:Lorcontinuation}) to imaginary external $c=1$ momenta is trivial\footnote{As noted before, the poles coming from the $c=1$ sphere four-point Virasoro conformal blocks are also independent of the external Liouville momenta $P_i$ for $i=1,\ldots,4$.}\textsuperscript{,}\footnote{Note that this is not a property of the $c=25$ structure constant $C(P_1,P_2,P)$, whose poles on the complex $P$-plane do depend on the values of $P_1$ and $P_2$.}\cite{Bautista:2019jau}.
Similarly, at the level of the torus one-point function (\ref{eq:torusCBdecompc1}), since the poles in the $\Phat$-integrand coming from the three-point coefficient $\Chat(P_{\rm ext},\Phat,\Phat)$ and from the torus one-point conformal blocks are independent of the external $c=1$ Liouville momenta $P_{\rm ext}$, the continuation to an imaginary value is trivial as well. 

Although clear from this discussion, in Appendix \ref{sec:consistency} we verify numerically that the sphere four-point function (\ref{eq:4ptLiouvc1}) and the torus one-point function (\ref{eq:torusCBdecompc1}) for imaginary external Liouville momenta continue to satisfy crossing symmetry and modular covariance, respectively. These results show that the continuation (\ref{eq:Lorcontinuation}) to Lorentzian energies is well-defined for any $n$-point function on all oriented Riemann surfaces in $c=1$ Liouville CFT. 

The spacetime interpretation of the string theory background (\ref{eq:wscft}), with the continuation to Lorentzian signature described above, is depicted in Figure \ref{fig:spacetime}. Our proposal is that worldsheet string perturbation theory computes a component of the \emph{cosmological wavefunction} defined as the overlap between the initial state of the universe (denoted by ``Big Bang") with an asymptotic out-state in the Hilbert space of perturbative string states, whose one-particle vertex operator representatives are given by (\ref{eq:vertexop}),
\ie
\Psi(\omega_1,\ldots,\omega_n) = \left\langle \, \omega_1,\ldots,\omega_n \mid \text{``Big Bang" } \right\rangle, ~~~~~ \ket{\omega_1,\ldots,\omega_n}\in\cal{H}^{\rm out}_{\rm string}.
\label{eq:cosmowavef}
\fe
The nature of the initial state of the universe is tied to the specific worldsheet CFT background under consideration. At present, it appears that string perturbation theory, starting from the background (\ref{eq:wscft}), is insufficient to explore different possibilities for this initial state. 
Our goal in this paper will be more modest, and we will study components of the cosmological wavefunction corresponding to the initial state of the universe implicitly defined by the specific worldsheet CFT (\ref{eq:wscft}).


\section{Cosmological wavefunction components in string perturbation theory}
\label{sec:cosmowavef}

In this section, we compute the three- and four-point components of the cosmological wavefunction at tree level in string perturbation theory.

\subsection{Three-point string diagram}
\label{sec:3ptwavef}

The first nontrivial cosmological wavefunction component computed in string perturbation theory is the tree-level three-point wavefunction component.
Fixing the position of the three vertex operators, the three-punctured sphere diagram has no remaining moduli and is evaluated in terms of the three-point coefficients of the matter sector reviewed in the previous section. We obtain that,
\ie
\Psi(\omega_1,\omega_2,\omega_3) &= \langle c\widetilde{c}\,\cV_{\omega_1}(0) c\widetilde{c}\,\cV_{\omega_2}(1) c\widetilde{c}\,\cV_{\omega_3}(\infty) \rangle \\
&= g_s^3 C_{S^2} \langle \Vhat_{-i\omega_1}(0) \Vhat_{-i\omega_2}(1) \Vhat_{-i\omega_3}(\infty) \rangle_{c=1\text{ Liouv.}} \langle V_{\omega_1}(0) V_{\omega_2}(1) V_{\omega_3}(\infty) \rangle_{c=25\text{ Liouv.}} \\
&= g_s^3 C_{S^2} \Chat(-i\omega_1,-i\omega_2,-i\omega_3) C(\omega_1,\omega_2,\omega_3) \\
&= 2^3 g_s^3 C_{S^2} \, \omega_1 \omega_2 \omega_3,
\label{eq:3ptWavef}
\fe
where $C_{S^2}$ is the normalization constant associated with the sphere topology. In time-independent string perturbation theory, perturbative unitarity (factorization of string scattering amplitudes) fixes the value of $C_{S^2}$ and is proportional to $g_s^{-2}$. In this paper, since at present we do not know the precise implications of unitarity on cosmological wavefunctions we will not fix $C_{S^2}$ and leave it as an undetermined constant. 




It is interesting to compare the three-point cosmological wavefunction component (\ref{eq:3ptWavef}) to the $1\to 2$ decay S-matrix element in $c=1$ string theory \cite{Balthazar:2017mxh},
\ie
\cS_{1\to 2} (\omega,\omega_1,\omega_2) = i\delta(\omega-\omega_1-\omega_2) \, \omega \omega_1 \omega_2,
\label{eq:c1_1to2smat}
\fe
which is also proportional to the product of the external energies. In $c=1$ string theory, whose matter sector consists of a time-like free boson and $c=25$ Liouville theory only, energy conservation is crucial in order for the DOZZ structure constant $C(\omega,\omega_1, \omega_2)$ to reduce to the simple product of the energies. In the two-dimensional string cosmology background the energies $\omega_i$ for $i=1,2,3$ are completely arbitrary, yet after multiplying the structure constants of both the $c=1$ and $c=25$ Liouville CFTs the wavefunction still simplifies drastically to the result (\ref{eq:3ptWavef}).

\subsection{Four-point string diagram}
\label{sec:4ptwavef}

Next, we compute the four-point cosmological wavefunction component in two-dimensional string cosmology. This is computed by the four-punctured sphere diagram which, after fixing the positions of three vertex operator insertions, has one remaining modulus corresponding to the position of the last vertex operator in the complex plane. 
The four-point wavefunction component takes the form
\ie
\Psi(\omega_1,\omega_2,\omega_3,\omega_4) \vphantom{\int} 
&= \int_{\bC}d^2z \,\langle c\widetilde{c}\,\cV_{\omega_4}(\infty) c\widetilde{c}\,\cV_{\omega_3}(1) \cV_{\omega_2}(z,\zbar) c\widetilde{c}\,\cV_{\omega_1}(0) \rangle \\
&= g_s^4 C_{S^2} \int_{\bC}d^2z \,\langle \Vhat_{-i\omega_4}(\infty) \Vhat_{-i\omega_3}(1) \Vhat_{-i\omega_2}(z,\zbar) \Vhat_{-i\omega_1}(0) \rangle_{c=1\text{ Liouv.}} \\
& ~~~~~~~~~~~~~~~~~ \times \langle V_{\omega_4}(\infty) V_{\omega_3}(1) V_{\omega_2}(z,\zbar) V_{\omega_1}(0) \rangle_{c=25\text{ Liouv.}} \vphantom{\int},
\label{eq:4ptWavef}
\fe
where the Liouville CFT four-point functions at $c=25$ and $c=1$ admit the Virasoro conformal block decompositions in (\ref{eq:4ptLiouvc25}) and (\ref{eq:4ptLiouvc1}), repectively. 

The moduli integral on the RHS of (\ref{eq:4ptWavef}) possibly diverges in regions of moduli space where the vertex operators collide, as is familiar in string perturbation theory. For instance, near $z=0$ the moduli integral of (\ref{eq:4ptWavef}) takes the form
\ie
\int_0 d^2z\int_{\cC}\frac{d\Phat}{2\pi} \int_0^\infty \frac{dP}{\pi} \Chat(-i\omega_1,-i\omega_2,\Phat) \Chat(-i\omega_3,-i\omega_4,\Phat) C(\omega_1,\omega_2,P) C(\omega_3,\omega_4,P) \, |z|^{-2+2P^2+2\Phat^2},
\label{eq:divergence}
\fe
where we have expanded both the $c=1$ and $c=25$ Virasoro conformal blocks to leading order in the small $z$ expansion using $\cF_c(h_i;h|z)\simeq z^{h-h_1-h_2}$, in our conventions. 
For small $z$, the integrals over the intermediate Liouville momenta $P$ and $\Phat$ are dominated by their values near $P=0$ and $\Phat=0$. Using Laplace's method to approximate these integrals, we obtain that the moduli integral as $z\to 0$ has the following behavior,
\ie
\frac{8}{\pi}\,\omega_1\omega_2\omega_3\omega_4\int_0 d^2z \,\frac{1}{|z|^2\left(\log|z|^2\right)^2},
\label{eq:nodiv}
\fe
and is therefore convergent. 
The limits as $z$ approaches 1 and infinity lead to same type of convergent behavior (\ref{eq:nodiv}), as can be easily seen by first using the crossing relations (\ref{eq:crossin13}) and (\ref{eq:crossin23}) to map each point to 0, respectively, and expanding the conformal blocks to leading order. 

In this paper, however, as detailed in Appendix \ref{sec:details4pt} we will follow closely the numerical integration strategy presented in \cite{Balthazar:2017mxh,Balthazar:2018qdv,Balthazar:2022atu}. In particular, we will switch the order of integrations and perform the moduli integral over $z$ first, and the integrals over the intermediate Liouville momenta $P$ and $\Phat$ last, which was a more convenient strategy of numerical integration in the computation of closed string scattering amplitudes in $c=1$ string theory \cite{Balthazar:2017mxh}. 
Following this order of integrations, and parametrizing the contour of integration $\cC$ of Figure \ref{fig:contourPhat} by $\Phat = p + i\epsilon$ for $p\in\bR$ and small $\epsilon > 0$, we observe from (\ref{eq:divergence}) that the four-point wavefunction component (\ref{eq:4ptWavef}) has a power divergence in $z$ whenever $P^2 + p^2 < \epsilon^2$. The region in the $(P,p)$ plane for which we have a divergence (and the divergence itself) can be made arbitrarily small as we take $\epsilon\to 0$. In this sense, the contour prescription depicted in Figure \ref{fig:contourPhat} serves as an ``$i\epsilon$ prescription" to regulate the cosmological wavefunction component (\ref{eq:4ptWavef}), when performing the $z$ moduli integral first. 
In practice, we will compute the four-point cosmological wavefunction component numerically for a small but fixed value of $\epsilon$, and hence we will need to properly regularize these divergences. We will do so following the procedure employed in \cite{Balthazar:2017mxh,Balthazar:2018qdv,Balthazar:2022atu} of counterterm subtraction to the moduli integral. The counterterms that remove the power divergences in the moduli $z$-integral are
\ie
R_s &= \iint\limits_{P^2+p^2\leq\epsilon^2} \frac{dP dp}{2\pi^2} \Chat(-i\omega_1,-i\omega_2,p+i\epsilon) \Chat(-i\omega_3,-i\omega_4,p+i\epsilon) C(\omega_1,\omega_2,P) C(\omega_3,\omega_4,P) \\
& ~~~~~~~~~~~~ \times|z|^{-2+2P^2+2p^2-2\epsilon^2+4ip\epsilon}, \vphantom{\int}\\
R_t &= \iint\limits_{P^2+p^2\leq\epsilon^2} \frac{dP dp}{2\pi^2} \Chat(-i\omega_2,-i\omega_3,p+i\epsilon) \Chat(-i\omega_1,-i\omega_4,p+i\epsilon) C(\omega_2,\omega_3,P) C(\omega_1,\omega_4,P) \\
& ~~~~~~~~~~~~ \times|z|^{-2+2P^2+2p^2-2\epsilon^2+4ip\epsilon}, \vphantom{\int}\\
R_u &= \iint\limits_{P^2+p^2\leq\epsilon^2} \frac{dP dp}{2\pi^2} \Chat(-i\omega_1,-i\omega_3,p+i\epsilon) \Chat(-i\omega_2,-i\omega_4,p+i\epsilon) C(\omega_1,\omega_3,P) C(\omega_2,\omega_4,P) \\
& ~~~~~~~~~~~~ \times|z|^{-2+2P^2+2p^2-2\epsilon^2+4ip\epsilon}. \vphantom{\int}
\label{eq:regulators}
\fe
The fully regularized four-point cosmological wavefunction in two-dimensional string cosmology is given by,
\ie
\Psi&(\omega_1,\omega_2,\omega_3,\omega_4) \vphantom{\Vhat}\\
&= g_s^4 C_{S^2} \int_{\bC}d^2z \left[\langle \Vhat_{-i\omega_4}(\infty) \Vhat_{-i\omega_3}(1) \Vhat_{-i\omega_2}(z,\zbar) \Vhat_{-i\omega_1}(0) \rangle_{c=1\text{ Liouv.}} \right. \\
& ~~~~~~~~~~~~~~~~~~~~~ \left. \times \langle V_{\omega_4}(\infty) V_{\omega_3}(1) V_{\omega_2}(z,\zbar) V_{\omega_1}(0) \rangle_{c=25\text{ Liouv.}} - R_s - R_t - R_u \vphantom{\Vhat}\right].
\label{eq:cosmo4ptreg}
\fe


In the direct numerical evaluation of the regularized cosmological wavefunction component (\ref{eq:cosmo4ptreg}) we will make the following choices for the outgoing energies of the asymptotic closed string states:
\ie
&({\rm i}) & ~~\omega_1 &= \omega_2 = \omega_3 = \omega_4 \equiv \omega, & ~~\omega \in [0,0.7], \vphantom{\frac{1}{1}}\\
&({\rm ii}) & \omega_1 &= \omega_2 = \omega_3 = \frac{1}{4},~~ \omega_4 \equiv \omega,  & \omega \in [0,0.7], \\
&({\rm iii}) & \omega_1 &= \omega_2 = \omega_3 = \frac{1}{3},~~ \omega_4 \equiv \omega,  & \omega \in [0,0.7], \\
&({\rm iv}) & \omega_1 &= \omega_2 = \omega_3 \equiv \omega, ~~ \omega_4 = \frac{1}{4}, &\omega \in [0,0.5],
\label{eq:choicenumerics}
\fe
and
\ie
&({\rm v}) & ~~\omega_1=\frac{1}{2},\, \omega_2=\frac{1}{3},\, \omega_3=\frac{1}{5},\, \omega_4 \equiv \omega,  & ~~~~~\omega \in [0,0.5], \vphantom{\frac{1}{1}} \vphantom{\int_0^0}\\
&({\rm vi}) & ~~\omega_1=\frac{1}{2},\, \omega_2=\frac{1}{3},\, \omega_3=\frac{3}{5},\, \omega_4 \equiv \omega,  & ~~~~~\omega \in [0,0.5] \vphantom{\frac{1}{1}} \vphantom{\int_0^0}.
\label{eq:choicegeneric}
\fe
The strategy we employ in the numerical calculation of the four-point cosmological wavefunction component follows \cite{Balthazar:2017mxh}.
We compute the four-point Virasoro conformal blocks numerically using Zamolodchikov's recursion relations \cite{Zamolodchikov:1985ie}, reviewed in Appendix \ref{sec:recursions}, truncated to a sufficiently high order in the elliptic nome $q$ series expansion, which is related to the cross-ratio $z$ by 
\ie
q(z) = \exp \left( -\pi\frac{K(1-z)}{K(z)} \right),
\label{eq:nomeq}
\fe
where $K(z)={}_{2}F_1(1/2,1/2;1|z)$ is the complete elliptic integral of the first kind. 
In order to obtain accurate numerical results with truncated conformal blocks, we first use crossing symmetry of the $c=1$ and $c=25$ Liouville four-point functions to reduce the moduli $z$-integration to a finite domain near the origin of the $q$-disc, or of the $z$-plane. 
The explicit expression for the four-point cosmological wavefunction component that we compute numerically is given by (\ref{eq:4ptfinal}). 
Further details of the calculation are given in Appendix \ref{sec:details4pt}.

\begin{figure}[h]
\centering

\begin{subfigure}[b]{0.495\textwidth}
\includegraphics[width=1\textwidth]{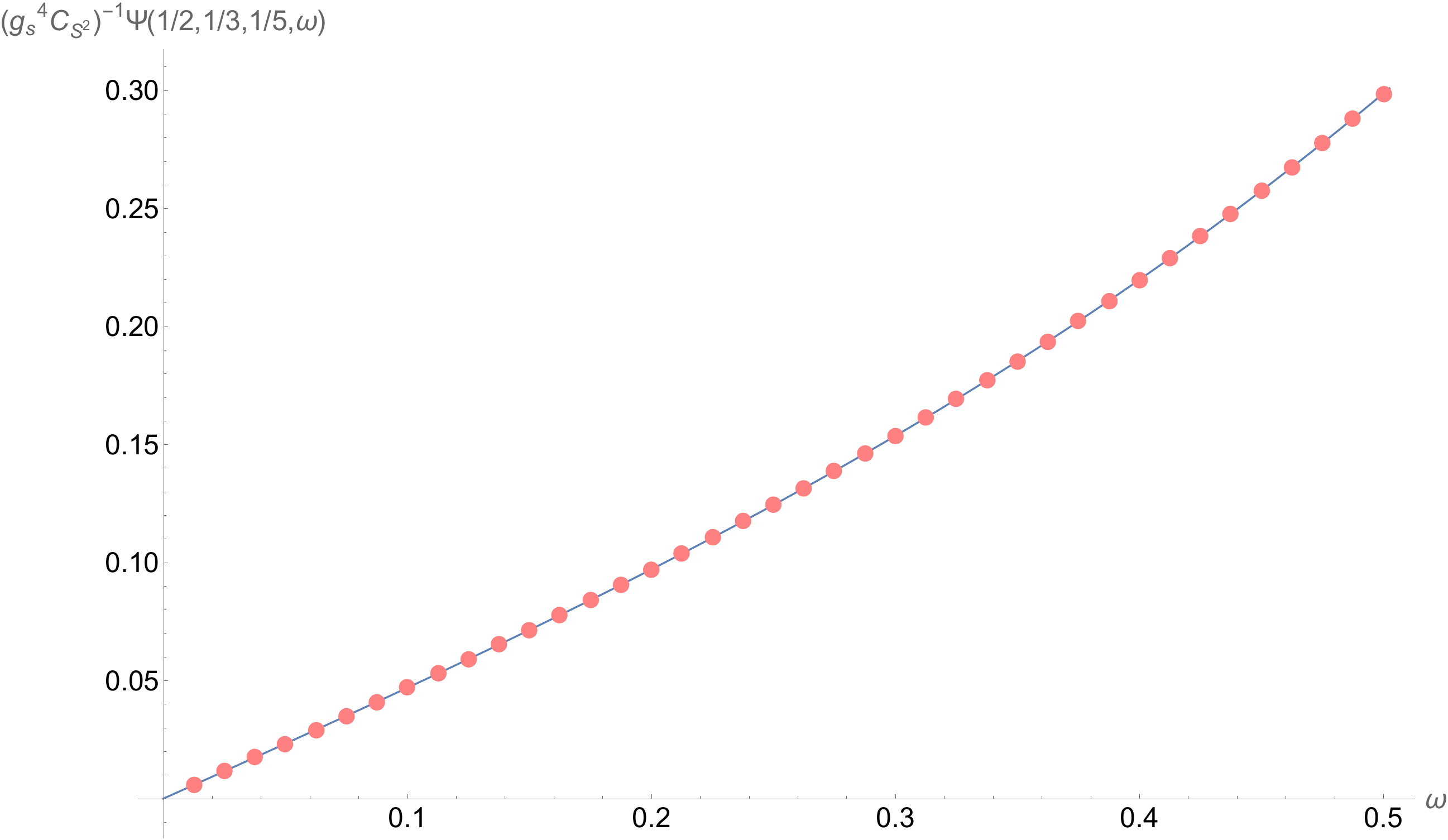}
\caption{$\Psi({1\over 2},{1\over 3},{1\over 5},\omega)$}
\end{subfigure}
\hfill
\begin{subfigure}[b]{0.495\textwidth}
\includegraphics[width=1\textwidth]{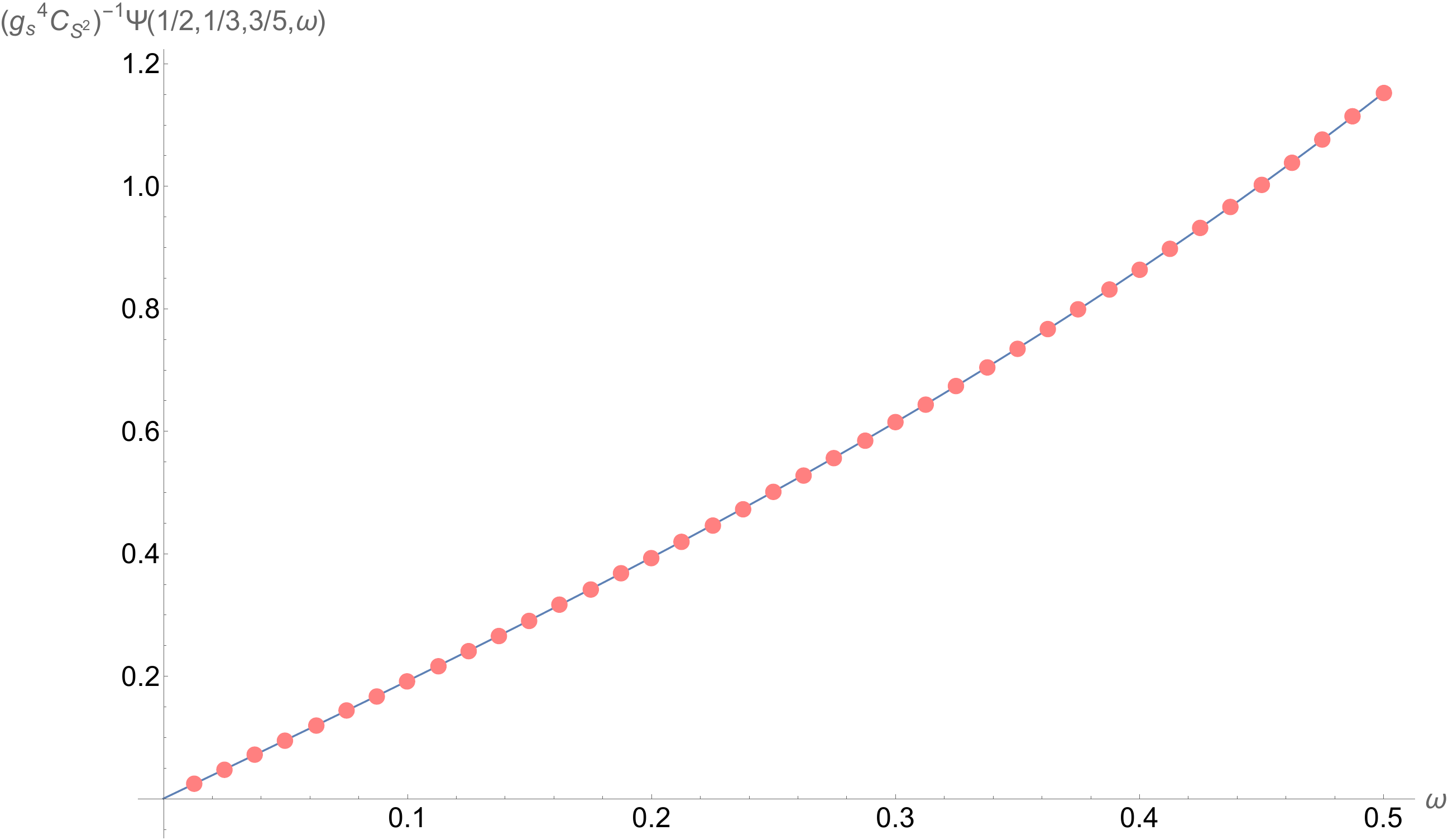}
\caption{$\Psi({1\over 2},{1\over 3},{3\over 5},\omega)$}
\end{subfigure}

\caption{Shown in dots are the numerical results for the regularized four-point cosmological wavefunction component in two-dimensional string cosmology, explicitly given by (\ref{eq:4ptfinal}), with the energy assignment (\ref{eq:choicegeneric}) for the outgoing asymptotic closed string states. The fit (\ref{eq:myfit}) is shown in the solid curve.}
\label{fig:resultsgenericw}
\end{figure}

The numerical results for the four-point cosmological wavefunction component (\ref{eq:cosmo4ptreg}) for the choices of outgoing closed string energies (\ref{eq:choicenumerics}) and (\ref{eq:choicegeneric}), following the strategy outlined in Appendix \ref{sec:details4pt}, are shown in Figures \ref{fig:results} and \ref{fig:resultsgenericw}.
We find to a high level of accuracy that a good fit to the numerical results is given by\footnote{Here the error is estimated by the discrepancy in the fit values from the four data sets (i)--(vi), not including the numerical error of the worldsheet computation itself.}
\ie
\Psi(\omega_1,\omega_2,\omega_3,\omega_4) ~=~ &g_s^4 C_{S^2} \, \omega_1 \omega_2 \omega_3 \omega_4 \left[ \alpha + \beta \left( \omega_1^2 + \omega_2^2 + \omega_3^2 + \omega_4^2 \right) \vphantom{\Vhat}\right], \vphantom{\int}\\
&\text{with } \alpha = 7.513 \pm 0.001, ~\beta = 15.99 \pm 0.01 \vphantom{\int}.
\label{eq:myfit}
\fe
\emph{Note added:} A more precise calculation has been performed in \cite{Collier:2023cyw}, resulting in the values $\alpha=8$ and $\beta=16$ in the conventions of the present paper.

Figures \ref{fig:results} and \ref{fig:resultsgenericw} show numerical results in pink dots and the fit (\ref{eq:myfit}) in a solid blue curve. The discrepancy between the numerical evaluation of (\ref{eq:4ptfinal}) and the fit (\ref{eq:myfit}) is at most $0.2\%$, but is typically much smaller than that. Further discussion of the possible sources of error in our numerical calculation of the regularized four-point cosmological wavefunction component is given in Appendix \ref{sec:details4pt}. 

The wavefunction component (\ref{eq:myfit}) is the main result of this paper. The fact that the three-point component (\ref{eq:3ptWavef}) and the four-point component (\ref{eq:myfit}) of the cosmological wavefunction take a very simple form is indicative of a dual description in terms of a matrix quantum mechanics, as is the case for the bosonic and type 0B two-dimensional string theories.
We hope that our result for the cosmological four-point wavefunction component (\ref{eq:myfit}) serves to identify the precise dual description of the cosmological background (\ref{eq:wscft}).

\begin{figure}[h]
\centering

\begin{subfigure}[b]{0.495\textwidth}
\includegraphics[width=1\textwidth]{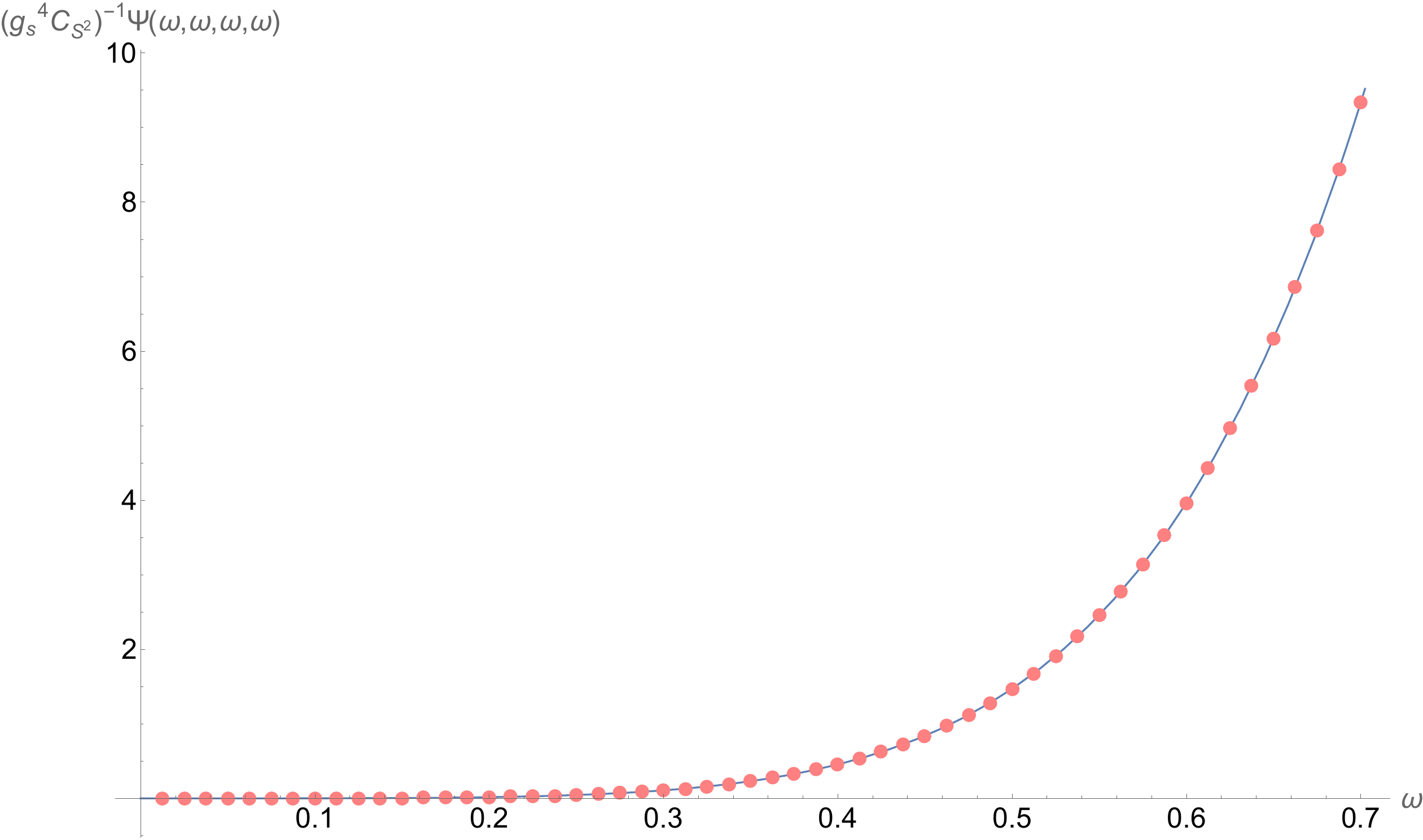}
\caption{$\Psi(\omega,\omega,\omega,\omega)$}
\end{subfigure}
\hfill
\begin{subfigure}[b]{0.495\textwidth}
\includegraphics[width=1\textwidth]{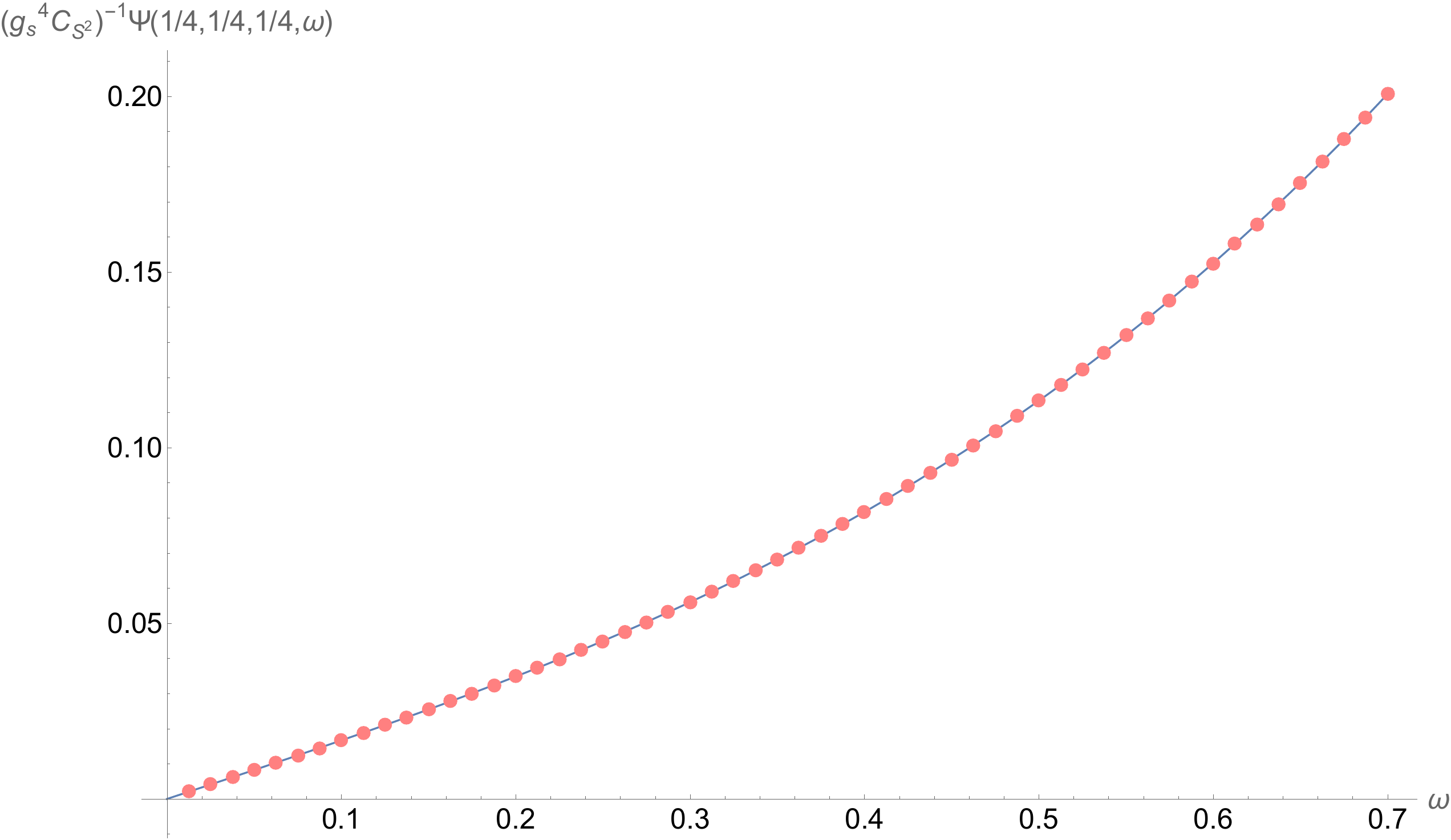}
\caption{$\Psi({1\over 4},{1\over 4},{1\over 4},\omega)$}
\end{subfigure}
~~\\
\begin{subfigure}[b]{0.495\textwidth}
\includegraphics[width=1\textwidth]{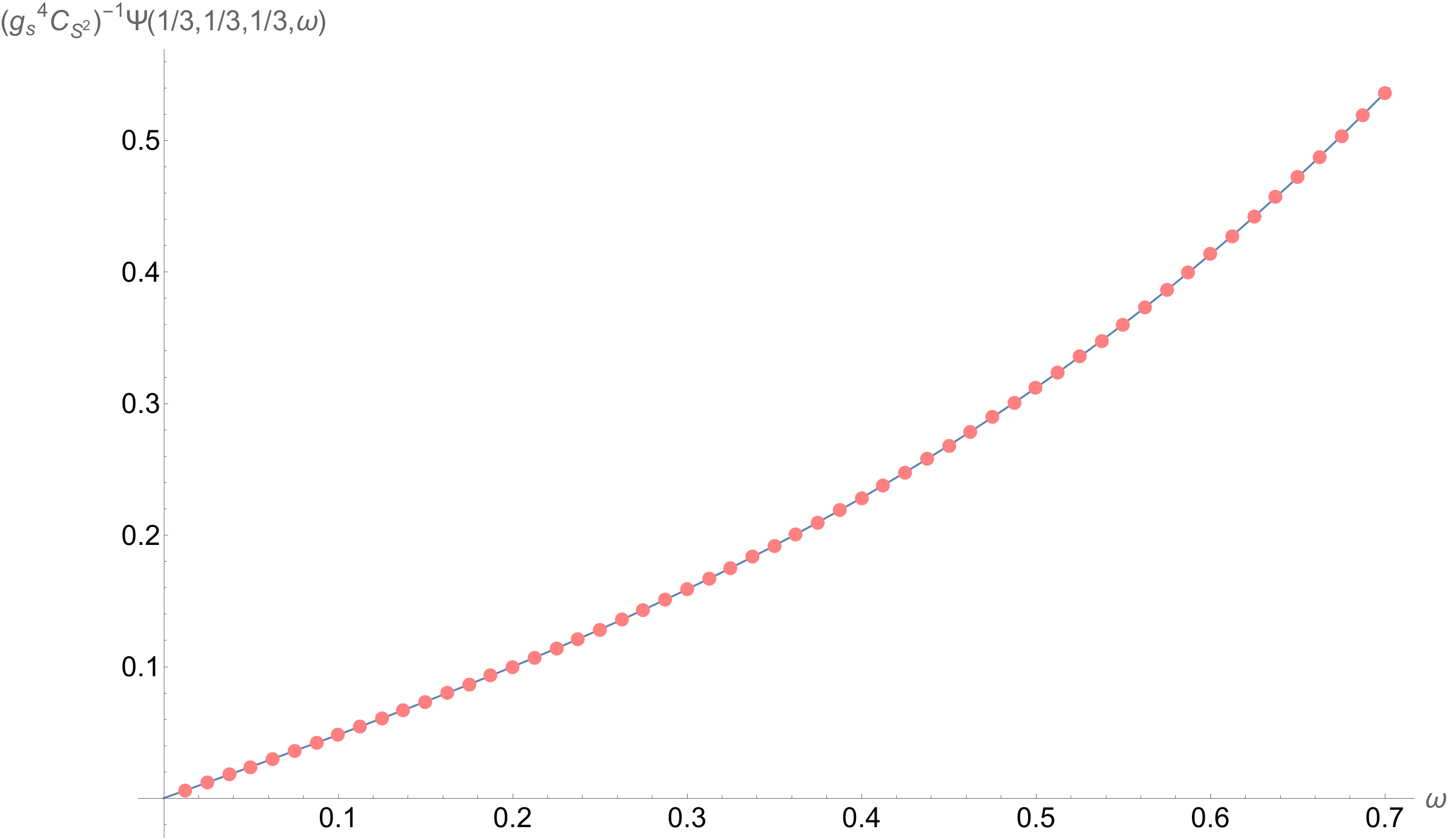}
\caption{$\Psi({1\over 3},{1\over 3},{1\over 3},\omega)$}
\end{subfigure}
\hfill
\begin{subfigure}[b]{0.495\textwidth}
\includegraphics[width=1\textwidth]{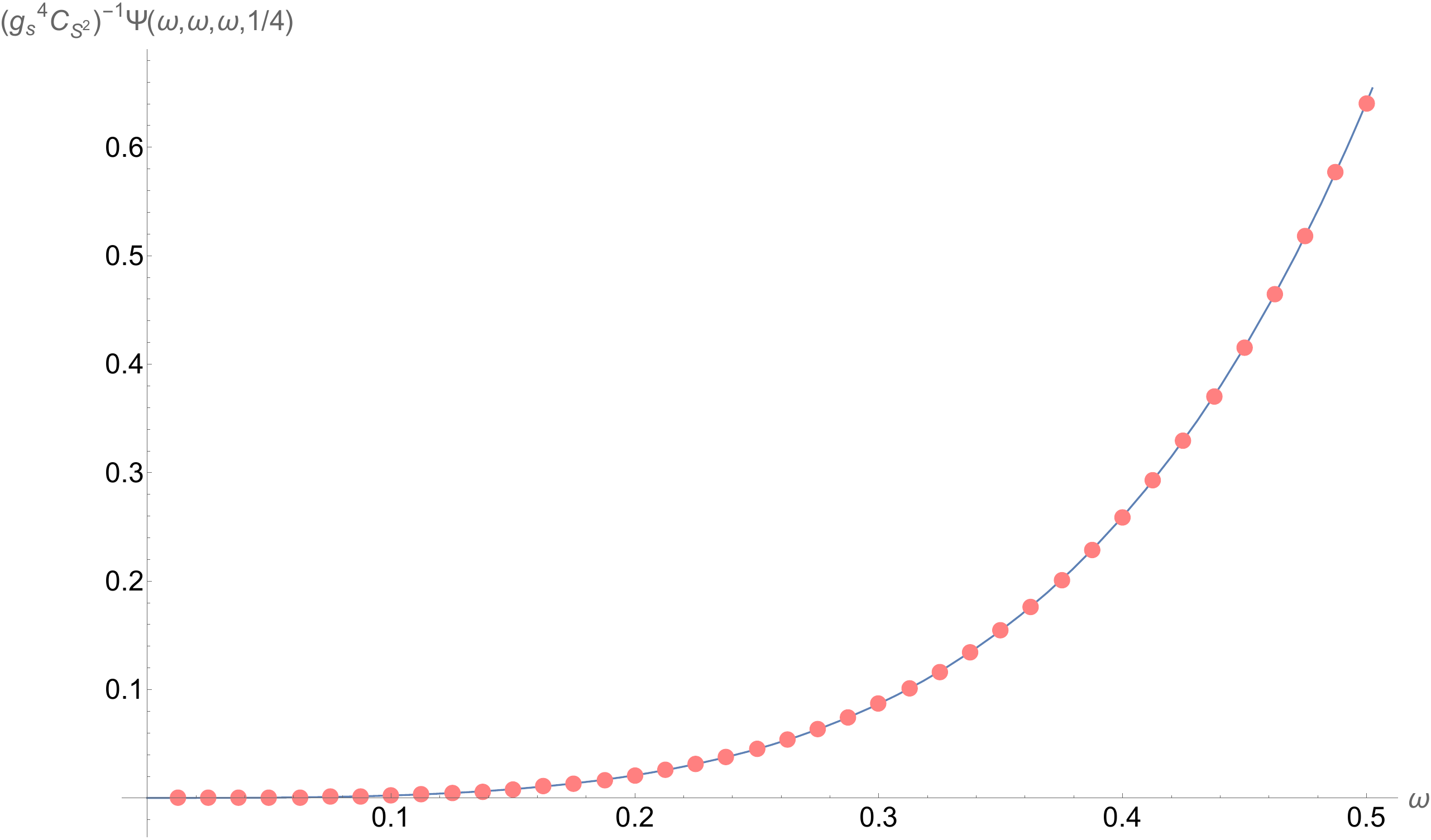}
\caption{$\Psi(\omega,\omega,\omega,{1\over 4})$}
\end{subfigure}

\caption{Shown in dots are the numerical results for the regularized four-point cosmological wavefunction component in two-dimensional string cosmology, explicitly given by (\ref{eq:4ptfinal}), with the energy assignment (\ref{eq:choicenumerics}) for the outgoing asymptotic closed string states. The fit (\ref{eq:myfit}) is shown in the solid curve.}
\label{fig:results}
\end{figure}


\section{Discussion}
\label{sec:discussion}

In this paper, we have studied in detail the time-dependent closed string background defined by the worldsheet CFT (\ref{eq:wscft}), after analytic continuation of the $c=1$ Liouville CFT to Lorentzian signature. 

In contrast to earlier work in the literature in the context of cosmological backgrounds in two-dimensional string theory\footnote{See for example \cite{CarneirodaCunha:2003mxy,Strominger:2003fn,Martinec:2003ka,Schomerus:2003vv,Hellerman:2007fc} and more recently \cite{Itzhaki:2021scf,Hashimoto:2022dro,Anninos:2021ene,Suzuki:2021zbe,Kapec:2020xaj}, which study various applications of ``time-like" Liouville theory.}, the main novelty of the present work is that the bootstrap solution of $c=1$ Liouville theory as a conformal field theory allowed for a description of the two-dimensional cosmological background that is exact in the string scale $\alpha'$ (which was set to $\alpha'=1$ throughout this paper). Hence, we were able to directly study components of the cosmological wavefunction using string perturbation theory. 
The main results of this paper are the explicit calculation of the tree-level three-point cosmological wavefunction component (\ref{eq:3ptWavef}) and the conjectural result (\ref{eq:myfit}), verified numerically, for the four-point wavefunction component.

The simplicity of the results (\ref{eq:3ptWavef}) and (\ref{eq:myfit}) is suggestive of the existence of a dual matrix quantum mechanics (MQM) description. 
Indeed, several time-dependent solutions of the dual MQM, which describe a time-dependent Fermi surface, have been studied in \cite{Alexandrov:2002fh,Alexandrov:2003uh,Karczmarek:2003pv} purely from the matrix side of the duality. 
One of the most interesting future directions is to identify the correct MQM dual to the string cosmological background described in this paper. 
This would provide an example of an exact holographic duality of a string theory in a cosmological background,  a subject for which we have limited understanding and examples are scarce.\footnote{In a different context, see \cite{Craps:2005wd,Craps:2006xq} that study time-dependent phenomena in a light-like linear dilaton background and its connection to BFSS matrix quantum mechanics \cite{Banks:1996vh}.} 

Another important question is to understand the nature of the initial state of the two-dimensional universe, which appears to be tied to the specific background (\ref{eq:wscft}). 
The classical worldsheet action (\ref{eq:timeLiouv}) of the time coordinate $\chi^0$ has an exponentially large potential in the infinite past. This suggests that the worldsheet of the string is prevented from going into the infinite past, and thus the initial state of the two-dimensional string universe looks empty. This property looks reminiscent of the Hartle-Hawking state.\footnote{We thank Daniel Jafferis for pointing this out.} 
Starting from the time-independent two-dimensional bosonic (``$c=1$ string theory") background, whose $c=1$ sector is a time-like free boson, a generic marginal deformation in general may correspond to a distinct initial state of the universe. Such deformations may not lead to an integrable worldsheet conformal field theory (and perhaps even to a local worldsheet theory). A special property of the initial state of the cosmological background studied in this paper is that it corresponds to a solvable worldsheet CFT. 

Another interesting direction is to understand more generally the implications of unitarity for string theory in cosmological spacetimes. In time-independent asymptotically flat spacetimes, unitarity is directly manifested as the unitarity of the string S-matrix. For example, string amplitudes in string perturbation theory satisfy unitary factorization into lower order amplitudes \cite{Polchinski:1998rq}. At present, we do not have a complete non-perturbative understanding of the implications of unitarity on cosmological wavefunctions (see \cite{Baumann:2022jpr,Flauger:2022hie} for a recent survey of research in this direction). 
Indeed, one motivation for the present work was to discover a simple example in string theory, hence ultra-violet complete, where the implications of unitarity for quantum field theory in cosmological spacetimes can be studied in detail.

\section*{Acknowledgements}

The author would like to thank Nima Arkani-Hamed, Bruno Balthazar, Victor Gorbenko, Daniel Jafferis, Austin Joyce, Juan Maldacena, Herman Verlinde, and Xi Yin for helpful discussions, Xi Yin for comments on a draft, and the organizers of the Abu Dhabi Meeting on Theoretical Physics for the hospitality during the course of this work. This research is supported in part by the Simons Collaboration Grant on the Nonperturbative Bootstrap and by the Future Faculty in the Physical Sciences Fellowship at Princeton University. 

This paper is dedicated to the memory of Manuel Victor Ruiz Ceh.

\appendix


\section{Consistency conditions in $c=1$ and $c=25$ Liouville CFT}
\label{sec:consistency}

In this section, we verify numerically the crossing symmetry of the sphere four-point function and the modular covariance of the torus one-point function in $c=1$ and $c=25$ Liouville theory\footnote{Modular covariance of the torus one-point function in $c\leq 1$ Liouville CFT was argued in \cite{Ribault:2015sxa} by using a special relation obeyed by the $c\leq 1$ three-point function coefficients (see their equation (3.3)). Here we test modular covariance numerically by directly integrating torus one-point Virasoro conformal blocks.}, which together imply the consistency of the CFTs on a general Riemann surface \cite{Moore:1988qv,Yin:2017yyn}. 

\subsection{Crossing symmetry of the sphere four-point function}

The first consistency condition on Liouville CFT is the crossing symmetry of the four-point function on the sphere. Our convention for the sphere four-point Virasoro conformal block decomposition of four scalar Virasoro primaries is such that
\ie
\langle \phi_4(z_4,\zbar_4) \phi_3(z_3,\zbar_3) \phi_2&(z_2,\zbar_2) \phi_1(z_1,\zbar_1) \rangle 
\vphantom{\sum}\\
=~ & |z_{24}|^{-4h_2} |z_{14}|^{2(h_2-h_1+h_3-h_4)} |z_{34}|^{2(h_2+h_1-h_3-h_4)} |z_{13}|^{2(h_4-h_1-h_2-h_3)} \vphantom{\sum}\\
&\times \sum_{h} C_{12h} C_{34h} \cF_c(h_i;h|z)\cF_c(h_i;h|\zbar), 
\label{eq:4pt}
\fe
where $z=\frac{z_{12}z_{34}}{z_{13}z_{24}}$ is the cross-ratio, $h_i$ for $i=1,\ldots,4$ are the external weights, and $h$ is the internal weight. 
For Liouville CFT at $c=25$ and at $c=1$, we define
\ie
G&(4321|z) \equiv \langle V_{P_4}(\infty) V_{P_3}(1) V_{P_2}(z,\zbar) V_{P_1}(0) \rangle_{c=25 \text{ Liouv.}} \vphantom{\int}\\
&= \int_0^\infty \frac{dP}{\pi} C(P_1,P_2,P) C(P_3,P_4,P) \cF_{c=25}(h_4,h_3,h_2,h_1;1+P^2|z)\cF_{c=25}(h_4,h_3,h_2,h_1;1+P^2|\zbar) ,\\
\widehat{G}&(4321|z) \equiv \langle \Vhat_{P_4}(\infty) \Vhat_{P_3}(1) \Vhat_{P_2}(z,\zbar) \Vhat_{P_1}(0) \rangle_{c=1 \text{ Liouv.}} \vphantom{\int}\\
&= \int_\cC \frac{d\Phat}{2\pi} \Chat(P_1,P_2,\Phat)\Chat(P_3,P_4,\Phat) \cF_{c=1}(\hhat_4,\hhat_3,\hhat_2,\hhat_1;\Phat^2|z)\cF_{c=1}(\hhat_4,\hhat_3,\hhat_2,\hhat_1;\Phat^2|\zbar),
\fe
where $\cC$ is the contour depicted in Figure \ref{fig:contourPhat}, $h_i=1+P_i^2$, and $\hhat_i=P_i^2$.
The crossing relation obtained by exchanging operators $1\leftrightarrow 3$ takes the form
\ie
G(4321|z) &= G(4123|1-z),\\
\widehat{G}(4321|z) &= \widehat{G}(4123|1-z).
\label{eq:crossin13}
\fe
This relation is obeyed at the level of individual Virasoro conformal blocks and hence it is not sensitive to the structure constants (\ref{eq:DOZZ1}) and (\ref{eq:DOZZ2}) of Liouville CFT at $c=25$ and $c=1$, respectively. 
On the other hand, the crossing relation obtained by exchanging operators $2\leftrightarrow 3$ takes the form
\ie
G(4321|z) &= |z|^{2(h_4-h_3-h_2-h_1)}G(4231|z^{-1}),\\
\widehat{G}(4321|z) &= |z|^{2(\hhat_4-\hhat_3-\hhat_2-\hhat_1)}\widehat{G}(4231|z^{-1}),
\label{eq:crossin23}
\fe
and is only satisfied after integration over the intermediate states against the three-point function coefficients.

Although clear from the analytic structure of the $\Phat$-integrand of $\widehat{G}(4321|z)$ as discussed in Section \ref{sec:c1Liouv}, we have also checked that the crossing relations (\ref{eq:crossin13}) and (\ref{eq:crossin23}) continue to hold for imaginary values of the external $c=1$ Liouville momenta, $P_i=-i\omega_i$ for $i=1,\ldots,4$.

\begin{figure}[h]
\centering

\begin{subfigure}[b]{0.495\textwidth}
\includegraphics[width=1\textwidth]{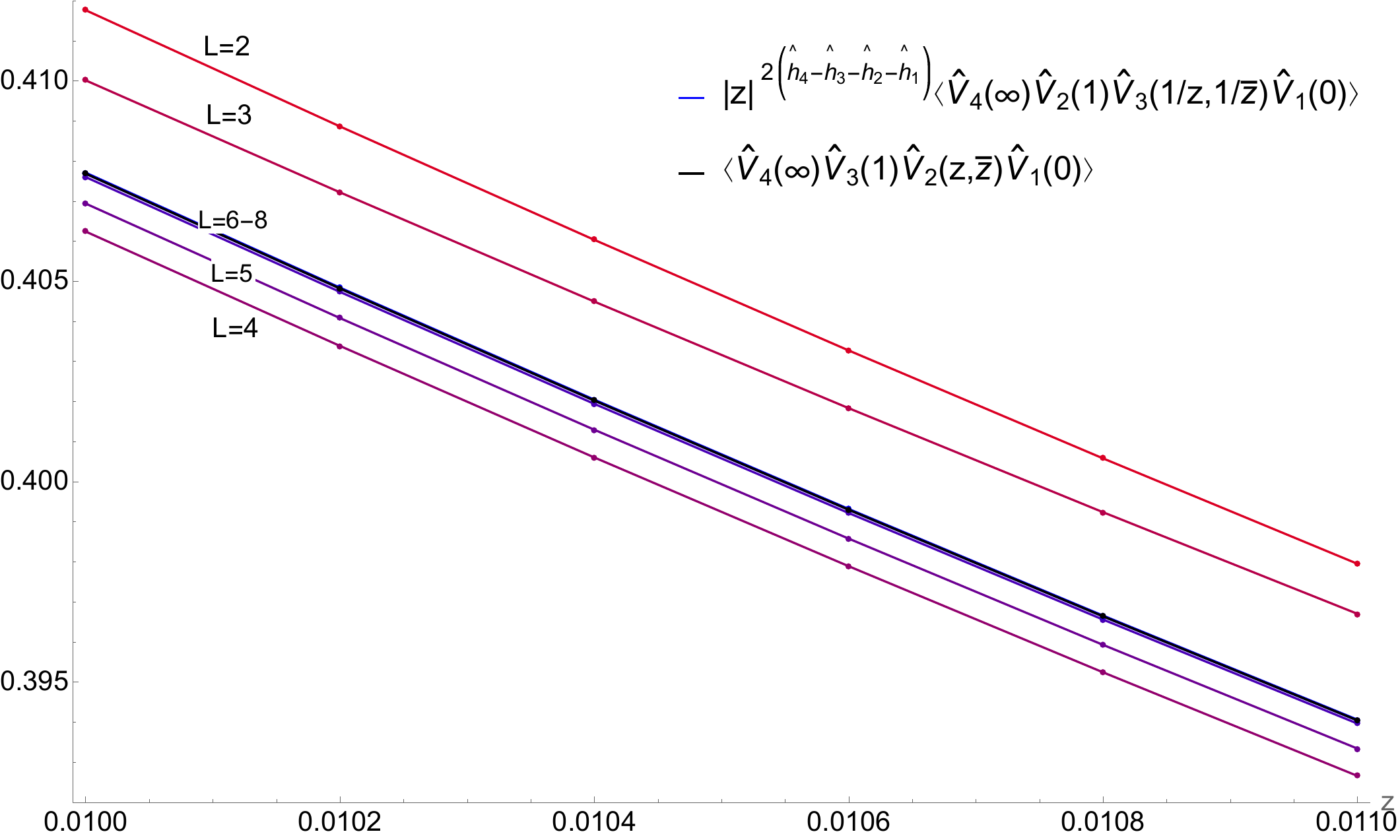}
\caption{$(P_1,P_2,P_3,P_4)=({1\over 3},{1\over 4},{1\over 5},{3\over 7})$}
\end{subfigure}
\hfill
\begin{subfigure}[b]{0.495\textwidth}
\includegraphics[width=1\textwidth]{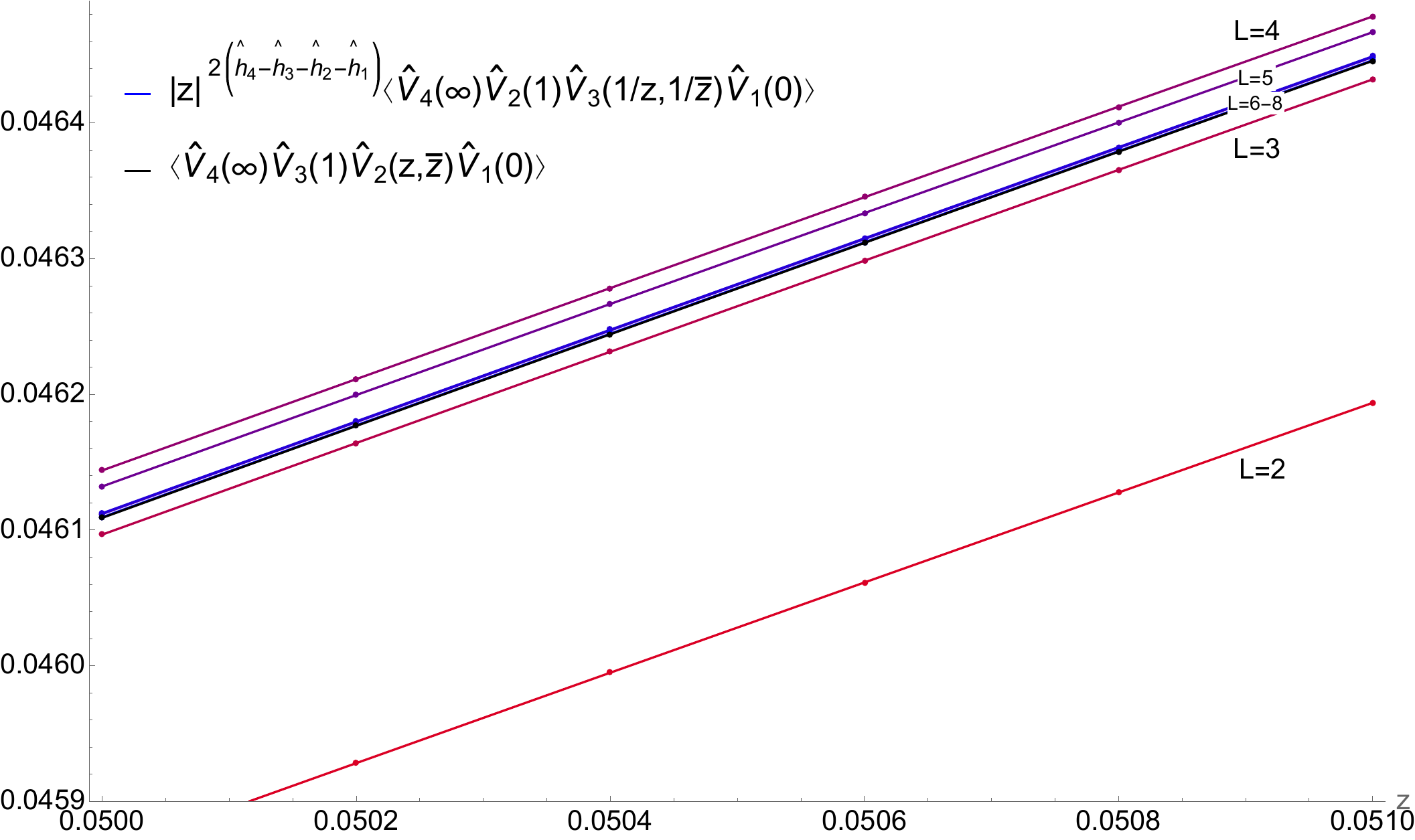}
\caption{$(P_1,P_2,P_3,P_4)=(-{i\over 3},-{i\over 4},-{i\over 5},-{3i\over 7})$}
\end{subfigure}

\caption{Numerical test of the crossing symmetry of $\langle \Vhat_{P_4}(\infty) \Vhat_{P_3}(1) \Vhat_{P_2}(z,\zbar) \Vhat_{P_1}(0) \rangle_{c=1 \text{ Liouv.}}$ for a range of real values of the cross-ratio $z$, at external $c=1$ Liouville momenta (i) $(P_1,P_2,P_3,P_4)=({1\over 3},{1\over 4},{1\over 5},{3\over 7})$ and (ii) $(P_1,P_2,P_3,P_4)=(-{i\over 3},-{i\over 4},-{i\over 5},-{3i\over 7})$. The crossed channel (RHS of (\ref{eq:crossin23})) computed with the sphere four-point conformal blocks truncated to order $q^{L}$ is shown with a color scheme from red to blue for increasing $L$ from 2 to 8. The direct channel (LHS of (\ref{eq:crossin23})) computed with conformal blocks truncated to order $q^8$ is shown in black. Data points are joined with straight lines for visualization.}
\label{fig:crossingplots}
\end{figure}

Figure \ref{fig:crossingplots} shows a sample verification of the crossing symmetry relation (\ref{eq:crossin23}) in $c=1$ Liouville CFT, for a purely real and a purely imaginary assignment of the external $c=1$ Liouville momenta $P_i$. For a range of positive real values for the cross-ratio $z$, the ``crossed" channel $13\to 24$ calculated with increasing  truncation order in the $q$ expansion of the conformal blocks (red to blue) can be seen to converge to the ``direct" channel $12\to 34$ result shown in black.

\subsection{Modular covariance of the torus one-point function}

The second consistency condition on Liouville CFT is the modular covariance of the one-point function of a Liouville vertex operator $V_{P_{\rm ext}}$ with external momentum $P_{\rm ext}$ on the torus with modulus $\tau$,
\ie
\left\langle V_{P_{\rm ext}}(0) \vphantom{\Vhat}\right\rangle^{T^2(\tau + 1)} = \left\langle V_{P_{\rm ext}}(0) \vphantom{\Vhat}\right\rangle^{T^2(\tau)},
\label{eq:modcovT}
\fe
and
\ie
\left\langle V_{P_{\rm ext}}(0) \vphantom{\Vhat}\right\rangle^{T^2(-{1\over \tau})} = (-i\tau)^{h_{\rm ext}} (i\overline{\tau})^{\widetilde{h}_{\rm ext}} \left\langle V_{P_{\rm ext}}(0) \vphantom{\Vhat}\right\rangle^{T^2(\tau)}, 
\label{eq:modcovS}
\fe
where $h_{\rm ext} = \widetilde{h}_{\rm ext}$ denote the conformal weights of the Liouville operator $V_{P_{\rm ext}}$. 
The first relation (\ref{eq:modcovT}) is a property of the torus one-point conformal blocks and thus insensitive to the Liouville three-point function coefficients. 
The second relation (\ref{eq:modcovS}) can be verified numerically in both $c=25$ and $c=1$ Liouville CFT by expanding the torus one-point function in terms of Virasoro conformal blocks as in (\ref{eq:torusCBdecompc25}) and (\ref{eq:torusCBdecompc1}), written again here for convenience
\ie
\left\langle V_{P_{\rm ext}}(0) \vphantom{\Vhat}\right\rangle^{T^2(\tau)}_{c=25 \text{ Liouv.}} &= \int_0^\infty \frac{dP}{\pi} C(P_{\rm ext},P,P) \cF_{c=25}(h_{\rm ext};1+P^2;q) \cF_{c=25}(h_{\rm ext};1+P^2;\overline{q}),\\
\left\langle \Vhat_{P_{\rm ext}}(0) \vphantom{\Vhat}\right\rangle^{T^2(\tau)}_{c=1 \text{ Liouv.}} &= \int_\cC \frac{d\Phat}{2\pi} \Chat(P_{\rm ext},\Phat,\Phat) \cF_{c=1}(\hhat_{\rm ext};\Phat^2|q) \cF_{c=1}(\hhat_{\rm ext};\Phat^2|\overline{q}),
\label{eq:torus1pts}
\fe
where $\cC$ is the contour depicted in Figure \ref{fig:contourPhat}. Using the recursion relations for the torus one-point Virasoro conformal blocks reviewed in Appendix \ref{sec:recursions}, we verified the relation (\ref{eq:modcovS}) for various values of the external Liouville momenta $P_{\rm ext}$ and of the torus modulus $\tau$ in both $c=1$ and $c=25$ Liouville CFT. Furthermore, and although clear from the analytic structure of the $\Phat$-integrand of the torus one-point function (\ref{eq:torusCBdecompc1}) as discussed in Section \ref{sec:c1Liouv}, we have also verified modular covariance (\ref{eq:modcovS}) for imaginary values of the external $c=1$ Liouville momenta, $P_{\rm ext}=-i\omega$. 

\begin{figure}[h]
\centering

\begin{subfigure}[b]{0.495\textwidth}
\includegraphics[width=1\textwidth]{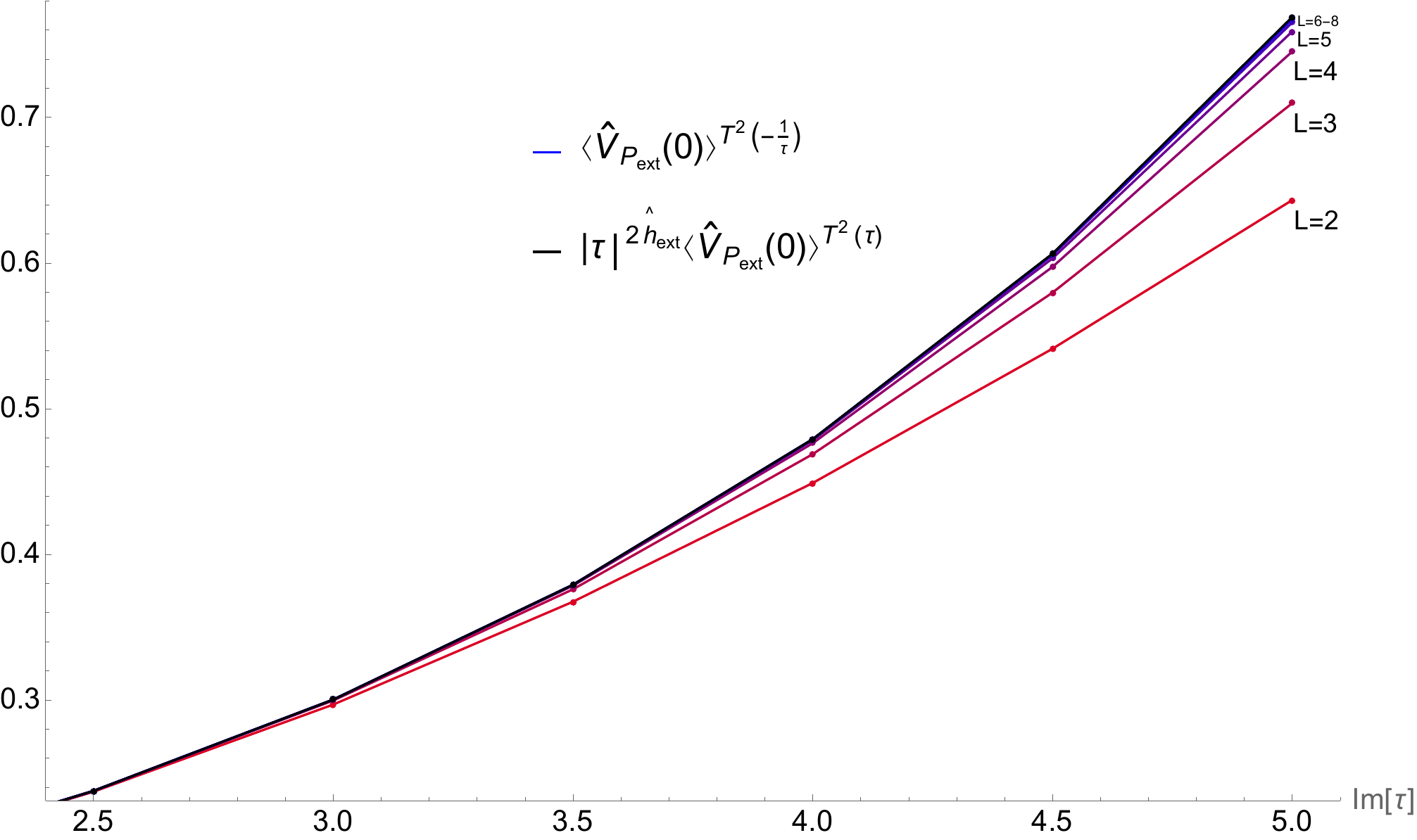}
\caption{$P_{\rm ext} = {1\over 3}$}
\end{subfigure}
\hfill
\begin{subfigure}[b]{0.495\textwidth}
\includegraphics[width=1\textwidth]{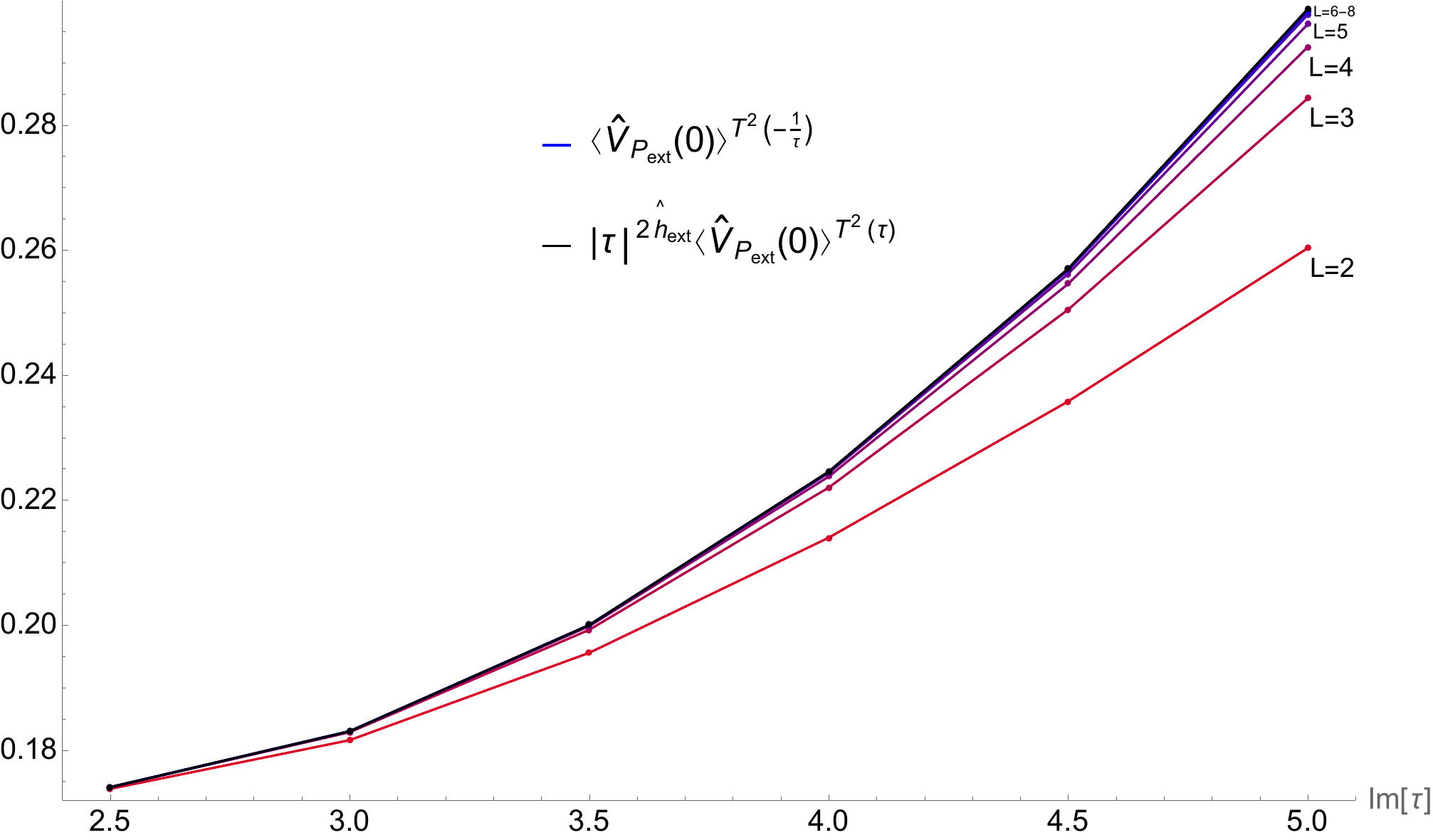}
\caption{$P_{\rm ext} = -{i\over 2}$}
\end{subfigure}

\caption{Numerical test of the modular covariance of $\left\langle \Vhat_{P_{\rm ext}}(0) \right\rangle^{T^2(\tau)}_{c=1 \text{ Liouv.}}$ for a range of values of $\tau$ with $\text{Re}\,\tau={1\over 3}$ and varying $\text{Im}\,\tau$, at external $c=1$ Liouville momenta (i) $P_{\rm ext}={1\over 3}$ and (ii) $P_{\rm ext}=-{i\over 2}$. The S-transformed channel (LHS of (\ref{eq:modcovS})) computed with the torus one-point conformal blocks truncated to order $q^{L}$ is shown with a color scheme from red to blue for increasing $L$ from 2 to 8. The direct channel (RHS of (\ref{eq:modcovS})) computed with conformal blocks truncated to order $q^8$ is shown in black. Data points are joined with straight lines for visualization.}
\label{fig:modcovplots}
\end{figure}

Figure \ref{fig:modcovplots} shows a sample verification of the modular covariance (\ref{eq:modcovS}) of the torus one-point function in $c=1$ Liouville CFT, for a real and a purely imaginary value of the external Liouville momenta. For a range of $\tau$ with increasing imaginary values, the ``S-transformed" channel (LHS of (\ref{eq:modcovS})) computed with increasing truncation order in the $q$ expansion of the conformal blocks (red to blue) can be seen to converge to the ``direct" channel result (RHS of (\ref{eq:modcovS})) shown in black.

\section{Recursive representations for the sphere 4-point and torus 1-point Virasoro conformal blocks}
\label{sec:recursions}

In this section we provide the recursion relations that we use to efficiently compute the sphere four-point  and the torus one-point Virasoro conformal blocks, originally derived in \cite{Zamolodchikov:1985ie} and in \cite{Hadasz:2009db,Cho:2017oxl}, respectively. 

We parametrize the central charge of the Virasoro algebra as $c=1+6Q^2$ with $Q=b+b^{-1}$, and the holomorphic weights of external primaries as
\ie
h_i = \frac{1}{4}\left( Q^2 - \lambda_i^2 \right).
\label{eq:lami}
\fe
We also define
\ie
h_{r,s} &= \frac{1}{4}\left( Q^2 - (rb+sb^{-1})^2 \right),\\
A_{r,s} &= \frac{1}{2}\underset{(p,q)\neq (0,0),(r,s)}{\prod_{p=1-r}^r \prod_{q=1-s}^s } \frac{1}{pb+qb^{-1}}.
\fe
The holomorphic sphere four-point Virasoro conformal block $\cF_c(h_i;h|z)$ with $i=1,\ldots,4$, defined in (\ref{eq:4pt}), can be expressed as \cite{Zamolodchikov:1985ie}
\ie
\cF_c(h_i;h|z) = (16q)^{h-Q^2/4} z^{Q^2/4-h_1-h_2} (1-z)^{Q^2/4-h_2-h_3} \theta_3(q)^{3Q^2 - 4(h_1+h_2+h_3+h_4)} \mathcal{H}_{c}(h_i;h|q),
\fe
where $\theta_3(q)$ is a Jacobi theta function, and the elliptic nome $q$ is related to the cross-ratio $z$ by
\ie
q(z) = \exp \left( -\pi\frac{K(1-z)}{K(z)} \right), ~~~~ \text{where } K(z)={}_{2}F_1(1/2,1/2;1|z).
\fe
The so-called ``elliptic conformal block" $\mathcal{H}_{c}(h_i;h|q)$ admits a power series expansion in $q$ and satisfies the following recursion relation,
\ie
\mathcal{H}_{c}(h_i;h|q) = 1 + \sum_{r,s \geq 1} (16q)^{rs} \frac{A_{r,s} P_{r,s}(h_1,h_2) P_{r,s}(h_4,h_3)}{h-h_{r,s}} \mathcal{H}_{c}(h_i;h\to h_{r,s} + rs|q),
\fe
where the ``fusion polynomials" $P_{r,s}$ are given by
\ie
P_{r,s}(h_1,h_2) = \prod_{p\in\{ 1-r,r-1,2 \}} \prod_{q\in \{ 1 - s, s - 1, 2\}} \frac{\lambda_1 + \lambda_2 + pb + qb^{-1}}{2}\frac{\lambda_1 - \lambda_2 + pb + qb^{-1}}{2},
\label{eq:fusion}
\fe
where the notation $p\in\{ 1-r,r-1,2 \}$ stands for $p$ ranging from $1-r$ to $r-1$ with step 2, and $\lambda_i$ are related to the external weights $h_i$ by (\ref{eq:lami}).

Similarly, the holomorphic tours one-point Virasoro conformal block $\cF_{c}(h_{\rm ext};h_{\rm int}|q)$ can be expressed as \cite{Hadasz:2009db}
\ie
\cF_{c}(h_{\rm ext};h_{\rm int}|q) = q^{h_{\rm int} - c/24} \left( \prod_{m=1}^{\infty} \frac{1}{1-q^m} \right) \mathcal{H}_c(h_{\rm ext};h_{\rm int}|q),
\fe
where now $q$ is related to the torus modulus $\tau$ by $q=e^{2\pi i \tau}$. The elliptic conformal block $\mathcal{H}_c(h_{\rm ext};h_{\rm int}|q)$ admits a power series expansion in $q$ and obeys the recursion relation,
\ie
\mathcal{H}_{c}(h_{\rm ext};h_{\rm int}|q) = 1 + \sum_{r,s \geq 1} q^{rs} \frac{A_{r,s} P_{r,s}(h_{\rm ext},h_{r,s}+rs) P_{r,s}(h_{\rm ext},h_{r,s})}{h_{\rm int}-h_{r,s}} \mathcal{H}_{c}(h_{\rm ext};h_{\rm int}\to h_{r,s} + rs|q),
\fe
where again the arguments of the fusion polynomials $P_{r,s}$ are related to the quantities $\lambda$ appearing in (\ref{eq:fusion}) by (\ref{eq:lami}). In this case, the product of the fusion polynomials may be written as
\ie
P&_{r,s}(h_{\rm ext},h_{r,s}+rs) P_{r,s}(h_{\rm ext},h_{r,s})\\
&=\prod_{k\in\{ 1,2r-1,2 \}} \prod_{l\in\{ 1,2s-1,2 \}} \frac{\lambda_{\rm ext} + kb + lb^{-1}}{2} \frac{\lambda_{\rm ext} - kb - lb^{-1}}{2} \frac{\lambda_{\rm ext} + kb - lb^{-1}}{2} \frac{\lambda_{\rm ext} - kb + lb^{-1}}{2}.
\fe

 A Mathematica notebook implementing these recursions relations is attached to this paper.


\section{Details of the numerical evaluation of the four-point cosmological wavefunction component}
\label{sec:details4pt}

In this section, we describe the strategy we employ for the numerical evaluation of the regularized four-point cosmological wavefunction component (\ref{eq:cosmo4ptreg}). It follows closely that of the four-point scattering amplitude in $c=1$ string theory of \cite{Balthazar:2017mxh}.

First, we make use of the crossing symmetry relations generated by (\ref{eq:crossin13}) and (\ref{eq:crossin23}) of the four-point correlation functions in Liouville CFT at $c=1$ and $c=25$ in order to reduce the moduli $z$-integration over the complex plane $\bC$ to a finite domain near $z=0$ as follows (see Appendix C.2 of \cite{Chang:2014jta}). 
We divide the complex plane $\bC$ into six regions: (I) ${\rm Re}\, z \leq 1/2, |1-z|\leq 1$, (II) $|z|\leq 1, |1-z|\geq 1$, (III) ${\rm Re}\, z \leq 1/2, |z|\geq 1$, (IV) ${\rm Re}\, z \geq 1/2, |z|\leq 1$, (V) $|1-z|\leq 1, |z|\geq 1$, and (VI) ${\rm Re}\, z \geq 1/2, |1-z|\geq 1$. 
Denote the transformation $z\to 1-z$, for which (\ref{eq:crossin13}) holds, by $T$ and the transformation $z\to z^{-1}$, for which (\ref{eq:crossin23}) holds, by $S$. The regions (II--VI) can be mapped to region I by the transformations $STS, TS, T, ST, S$, respectively. 
After these transformations, and if we parametrize the contour of integration $\cC$ over the $c=1$ intermediate Liouville momenta by $\Phat=p+i\epsilon$ with $p\in\bR$ and $\epsilon>0$, the regularized cosmological wavefunction component (\ref{eq:cosmo4ptreg}) can be written as
\ie
(&g_s^4 C_{S^2})^{-1}\,\Psi(\omega_1,\omega_2,\omega_3,\omega_4) \vphantom{\int}\\
& = \int\limits_{\text{region I}}d^2z \left\{\left[ \int_0^{\infty} \frac{dP}{\pi} \int_{-\infty}^\infty \frac{dp}{2\pi} C(\omega_1,\omega_2,P)C(\omega_3,\omega_4,P) \Chat(-i\omega_1,-i\omega_2,p+i\epsilon)\Chat(-i\omega_3,-i\omega_4,p+i\epsilon) \right.\right.\\
& ~~~~~~~~~~~~~~~~~~~~~~~~~~~~ \times\cF_{c=25}(h_4,h_3,h_2,h_1;1+P^2|z)\cF_{c=25}(h_4,h_3,h_2,h_1;1+P^2|\zbar), \vphantom{\int}\\
& ~~~~~~~~~~~~~~~~~~~~~~~~~~~~ \times \cF_{c=1}(\hhat_4,\hhat_3,\hhat_2,\hhat_1;(p+i\epsilon)^2|z)\cF_{c=1}(\hhat_4,\hhat_3,\hhat_2,\hhat_1;(p+i\epsilon)^2|\zbar) \vphantom{\int}\\
& ~~~~~~~~~~ ~~~ - \frac{1}{2} \iint\limits_{P^2+p^2\leq\epsilon^2} \frac{dP dp}{2\pi^2} C(\omega_1,\omega_2,P)C(\omega_3,\omega_4,P) \Chat(-i\omega_1,-i\omega_2,p+i\epsilon)\Chat(-i\omega_3,-i\omega_4,p+i\epsilon) \\
& ~~~~~~~~~~ ~~~~~~~~~~~~~~~ ~~~ \times \left( |z|^{-2+2P^2+2(p+i\epsilon)^2} + |z|^{-2+2P^2+2(p+i\epsilon)^2}|1-z|^{-2-2P^2-2(p+i\epsilon)^2} \right. \vphantom{\int}\\
& ~~~~~~~~~~ ~~~~~~~~~~~~~~~ ~~~~~~~~ + |z|^{-2-2P^2-2(p+i\epsilon)^2}|1-z|^{-2+2P^2+2(p+i\epsilon)^2} + |1-z|^{-2+2P^2+2(p+i\epsilon)^2} \vphantom{\int}\\
& ~~~~~~~~~~ ~~~~~~~~~~~~~~~ ~~~~~~~~ \left.\left. + |1-z|^{-2-2P^2-(p+i\epsilon)^2} + |z|^{-2-2P^2-2(p+i\epsilon)^2} \right) \vphantom{\int }\right] \\
& \left. ~~~~~~~~~~ + \left[  \text{ other 5 permutations of \{123\} } \vphantom{\Vhat}\right]  \vphantom{\int}\right\}.
\label{eq:4ptfinal}
\fe

For the numerical evaluation of (\ref{eq:4ptfinal}), we further cut out a disc of small radius $\delta>0$ around $z=0$, $D_\delta\equiv\{ |z|\leq \delta \}$. Inside the region $D_\delta \cap {\rm I}$, we truncate the Virasoro conformal blocks to next-to-leading order in $z$ and perform the $z$-integral analytically, including the regulator counterterms. Outside this cut-out disc, for $\overline{D_\delta}\cap{\rm I}$ we perform the $z$-integral numerically with the conformal blocks truncated to a sufficiently high order in the $q$-recursion expansion. The results shown in Figures \ref{fig:results} and \ref{fig:resultsgenericw} were computed with conformal blocks  truncated to order $q^8$, and with a choice of $\delta=10^{-2}$.

\begin{figure}[h]
\centering

\begin{subfigure}[b]{0.495\textwidth}
\includegraphics[width=1\textwidth]{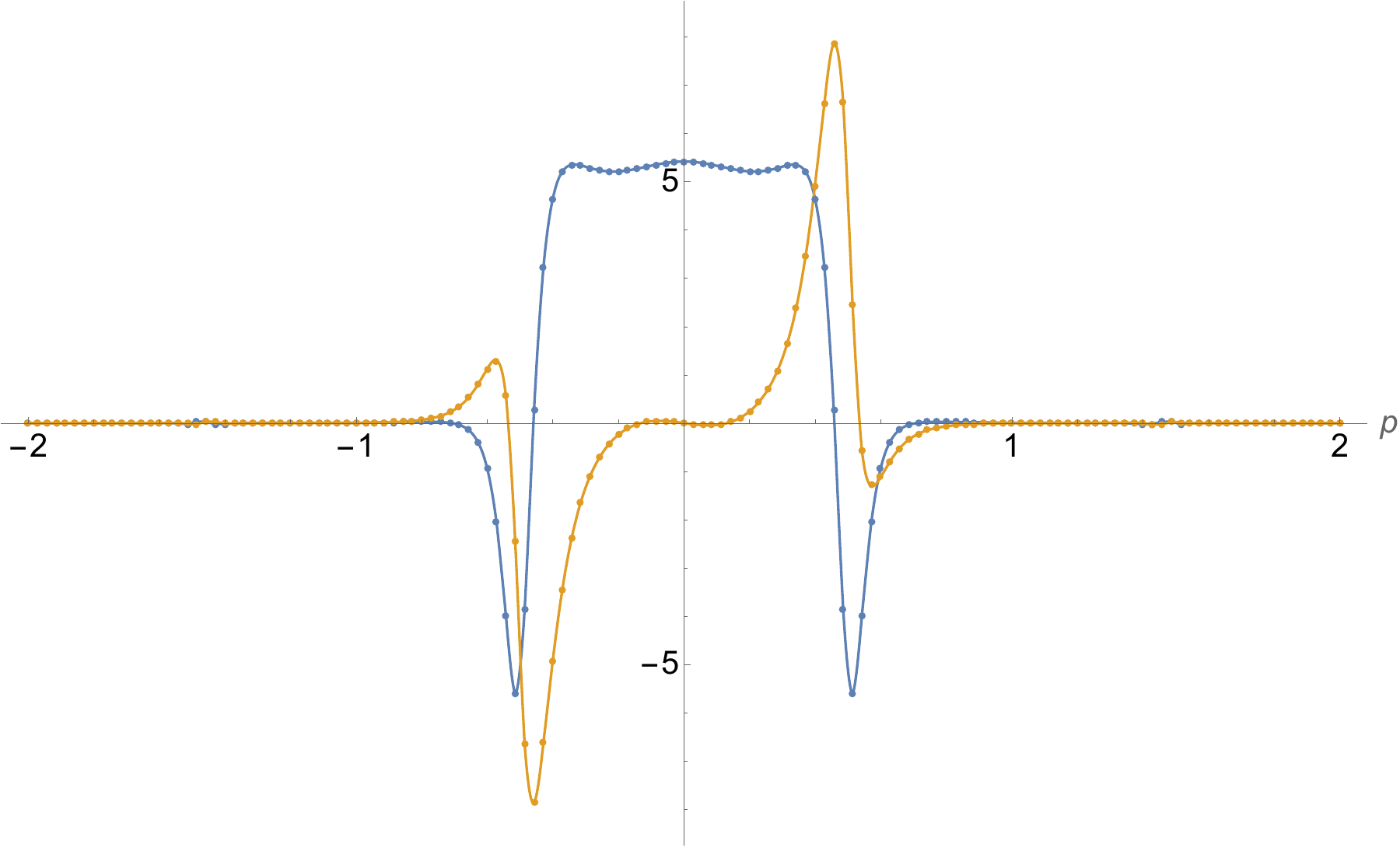}
\caption{$\Phat$-integrand}
\end{subfigure}
\hfill
\begin{subfigure}[b]{0.495\textwidth}
\includegraphics[width=1\textwidth]{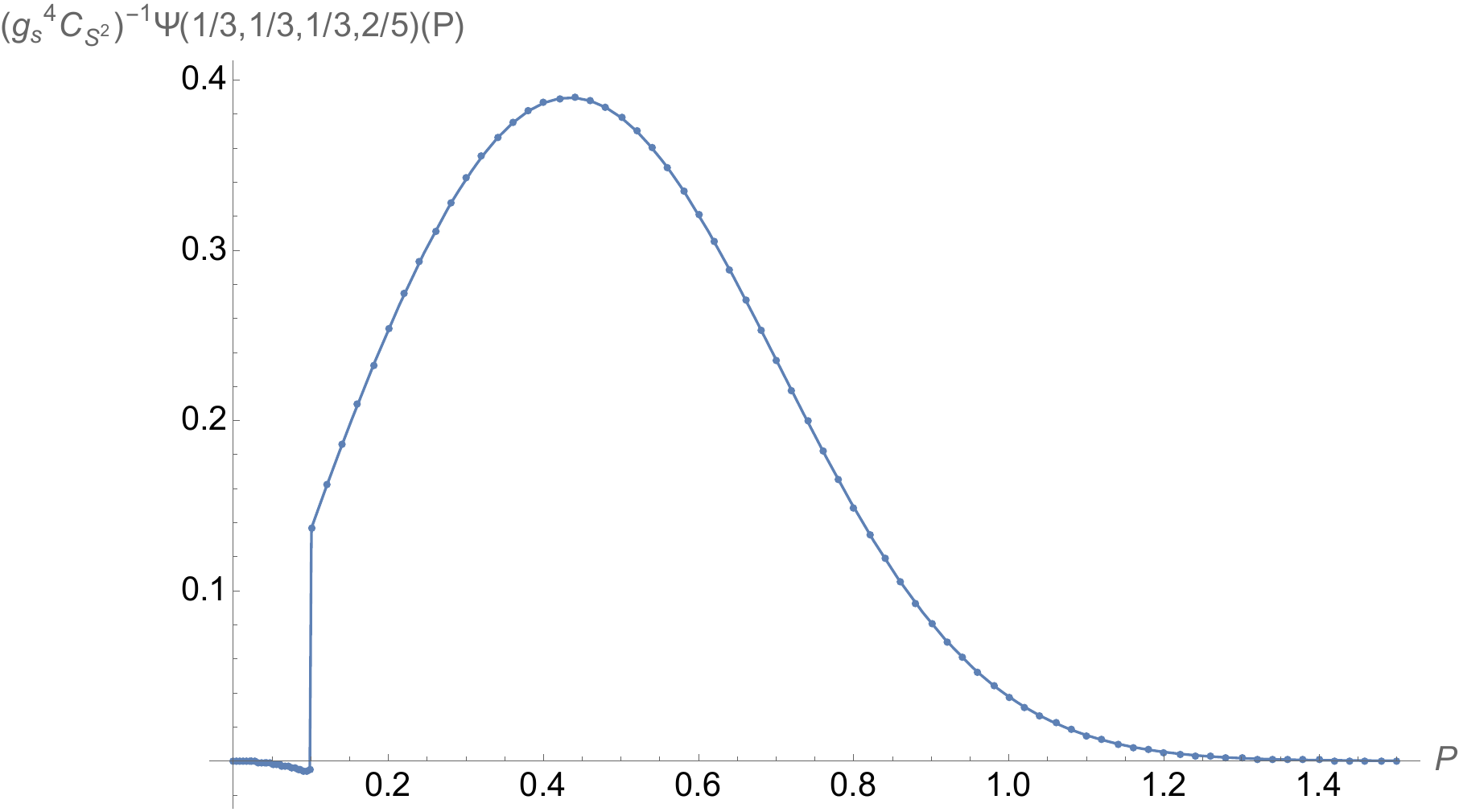}
\caption{$P$-integrand}
\end{subfigure}

\caption{Sample plots at intermediate stages in the numerical calculation of the regularized four-point cosmological wavefunction component (\ref{eq:4ptfinal}) for the choice of $\omega_1=\omega_2=\omega_3={1\over 3}$ and $\omega_4={2\over 5}$. (i) shows the $\Phat$-integrand of (\ref{eq:4ptfinal}) for a fixed value of $P={1\over 2}$ after having performed the $z$-integration over region (I) with the strategy outlined above with $\delta=10^{-2}$. (ii) shows the $P$-integrand after having performed both the $z$ moduli integration and the $\Phat$ integration over the contour $\cC$ parametrized by $\Phat=p+i\epsilon$ with $\epsilon=10^{-1}$.}
\label{fig:sample4pt}
\end{figure}

As mentioned in the main text, in the numerical evaluation of (\ref{eq:4ptfinal}) we switch the order of integrations, following \cite{Balthazar:2017mxh}.  We first perform the moduli $z$-integral over region (I) with the strategy described above for a fixed value of the $c=1$ and $c=25$ intermediate Liouville momenta $\Phat$ and $P$. Second, we perform the $c=1$ Liouville $\Phat$-integral over the contour $\cC$ parametrized by $\Phat=p+i\epsilon$; the results reported in Figures \ref{fig:results} and \ref{fig:resultsgenericw} were computed with $\epsilon=10^{-1}$. Lastly, we perform the $c=25$ Liouville $P$-integral. Sample plots of (i) the $\Phat$-integrand and (ii) the $P$-integrand are shown in Figure \ref{fig:sample4pt}. The discontinuity in the $P$-integrand visible in Figure \ref{fig:sample4pt} (ii) is due to the regulator counterterms that are only nonzero for $P\leq \epsilon=10^{-1}$.

As we approach the limit of $c\to 25$ and $c\to 1$, the recursion relations for the Virasoro conformal blocks involve delicate cancelations. In order to avoid loss of precision when computing the conformal blocks recursively, we work with sufficiently close values (but not exactly equal) to the Liouville central charges. For the results reported in Figures \ref{fig:results} and \ref{fig:resultsgenericw}, we set $c=0.995$ and $c=25.00001$. In fact, this appears to be the largest source of numerical error in our calculation of (\ref{eq:4ptfinal}). 

To a lesser extent, another source of error is the numerical integration in the cross-ratio $z$ over region (I), and the numerical interpolation and subsequent integration over the Liouville intermediate momenta $P$ and $\Phat$, as in the sample plots in Figure \ref{fig:sample4pt}. 
In addition, for larger values of the outgoing closed strings energies $\omega_i$, both Liouville momenta integrations have wider non-vanishing support. The results presented in Figures \ref{fig:results} and \ref{fig:resultsgenericw} were calculated by integrating up to a maximum value of $|p|=2$ and $P=2.5$, which could be another source of small error for larger values of the outgoing energies $\omega_i$.


\bibliographystyle{JHEP}
\bibliography{2dStringCosmology_refs}

\providecommand{\href}[2]{#2}\begingroup\raggedright\begin{thebibliography}{10}

\bibitem{Klebanov:1991qa}
I.~R. Klebanov, {\it {String theory in two-dimensions}},  in {\em {Spring
  School on String Theory and Quantum Gravity (to be followed by Workshop)
  Trieste, Italy, April 15-23, 1991}}, pp.~30--101, 1991.
\newblock \href{http://arxiv.org/abs/hep-th/9108019}{{\tt hep-th/9108019}}.

\bibitem{Ginsparg:1993is}
P.~H. Ginsparg and G.~W. Moore, {\it {Lectures on 2-D gravity and 2-D string
  theory}},  in {\em {Theoretical Advanced Study Institute (TASI 92): From
  Black Holes and Strings to Particles Boulder, Colorado, June 3-28, 1992}},
  pp.~277--469, 1993.
\newblock \href{http://arxiv.org/abs/hep-th/9304011}{{\tt hep-th/9304011}}.
\newblock [,277(1993)].

\bibitem{Jevicki:1993qn}
A.~Jevicki, {\it {Development in 2-d string theory}},  in {\em {Workshop on
  String Theory, Gauge Theory and Quantum Gravity Trieste, Italy, April 28-29,
  1993}}, pp.~96--140, 1993.
\newblock \href{http://arxiv.org/abs/hep-th/9309115}{{\tt hep-th/9309115}}.

\bibitem{Polchinski:1994mb}
J.~Polchinski, {\it {What is string theory?}},  in {\em {NATO Advanced Study
  Institute: Les Houches Summer School, Session 62: Fluctuating Geometries in
  Statistical Mechanics and Field Theory Les Houches, France, August
  2-September 9, 1994}}, 1994.
\newblock \href{http://arxiv.org/abs/hep-th/9411028}{{\tt hep-th/9411028}}.

\bibitem{Martinec:2004td}
E.~J. Martinec, {\it {Matrix models and 2D string theory}},  in {\em {9th
  Frontiers of Mathematical Physics Summer School on Strings, Gravity and
  Cosmology Vancouver, Canada, August 2-13, 2004}}, pp.~403--457, 2004.
\newblock \href{http://arxiv.org/abs/hep-th/0410136}{{\tt hep-th/0410136}}.
\newblock [,403(2004)].

\bibitem{Nakayama:2004vk}
Y.~Nakayama, {\it {Liouville field theory: A Decade after the revolution}},
  {\em Int. J. Mod. Phys. A} {\bf 19} (2004) 2771--2930,
  [\href{http://arxiv.org/abs/hep-th/0402009}{{\tt hep-th/0402009}}].

\bibitem{Brezin:1989ss}
E.~Brezin, V.~A. Kazakov, and A.~B. Zamolodchikov, {\it {Scaling Violation in a
  Field Theory of Closed Strings in One Physical Dimension}},  {\em Nucl. Phys.
  B} {\bf 338} (1990) 673--688.

\bibitem{Gross:1990ay}
D.~J. Gross and N.~Miljkovic, {\it {A Nonperturbative Solution of $D=1$ String
  Theory}},  {\em Phys. Lett. B} {\bf 238} (1990) 217--223.

\bibitem{Ginsparg:1990as}
P.~H. Ginsparg and J.~Zinn-Justin, {\it {2-d GRAVITY + 1-d MATTER}},  {\em
  Phys. Lett. B} {\bf 240} (1990) 333--340.

\bibitem{Douglas:2003up}
M.~R. Douglas, I.~R. Klebanov, D.~Kutasov, J.~M. Maldacena, E.~J. Martinec, and
  N.~Seiberg, {\it {A New hat for the c=1 matrix model}},  7, 2003.
\newblock \href{http://arxiv.org/abs/hep-th/0307195}{{\tt hep-th/0307195}}.

\bibitem{Takayanagi:2003sm}
T.~Takayanagi and N.~Toumbas, {\it {A Matrix model dual of type 0B string
  theory in two-dimensions}},  {\em JHEP} {\bf 07} (2003) 064,
  [\href{http://arxiv.org/abs/hep-th/0307083}{{\tt hep-th/0307083}}].

\bibitem{Balthazar:2019rnh}
B.~Balthazar, V.~A. Rodriguez, and X.~Yin, {\it {ZZ Instantons and the
  Non-Perturbative Dual of $c =$ 1 String Theory}},
  \href{http://arxiv.org/abs/1907.07688}{{\tt arXiv:1907.07688}}.

\bibitem{Balthazar:2019ypi}
B.~Balthazar, V.~A. Rodriguez, and X.~Yin, {\it {Multi-Instanton Calculus in $c
  = 1$ String Theory}},  \href{http://arxiv.org/abs/1912.07170}{{\tt
  arXiv:1912.07170}}.

\bibitem{Sen:2019qqg}
A.~Sen, {\it {Fixing an Ambiguity in Two Dimensional String Theory Using String
  Field Theory}},  {\em JHEP} {\bf 03} (2020) 005,
  [\href{http://arxiv.org/abs/1908.02782}{{\tt arXiv:1908.02782}}].

\bibitem{Sen:2020oqr}
A.~Sen, {\it {Divergent $\Longrightarrow$ complex amplitudes in two dimensional
  string theory}},  {\em JHEP} {\bf 02} (2021) 086,
  [\href{http://arxiv.org/abs/2003.12076}{{\tt arXiv:2003.12076}}].

\bibitem{Sen:2020ruy}
A.~Sen, {\it {Cutkosky Rules and Unitarity (Violation) in D-instanton
  Amplitudes}},  \href{http://arxiv.org/abs/2012.00041}{{\tt
  arXiv:2012.00041}}.

\bibitem{Sen:2020eck}
A.~Sen, {\it {D-instantons, string field theory and two dimensional string
  theory}},  {\em JHEP} {\bf 11} (2021) 061,
  [\href{http://arxiv.org/abs/2012.11624}{{\tt arXiv:2012.11624}}].

\bibitem{Sen:2021qdk}
A.~Sen, {\it {Normalization of D-instanton amplitudes}},  {\em JHEP} {\bf 11}
  (2021) 077, [\href{http://arxiv.org/abs/2101.08566}{{\tt arXiv:2101.08566}}].

\bibitem{DeWolfe:2003qf}
O.~DeWolfe, R.~Roiban, M.~Spradlin, A.~Volovich, and J.~Walcher, {\it {On the S
  matrix of type 0 string theory}},  {\em JHEP} {\bf 11} (2003) 012,
  [\href{http://arxiv.org/abs/hep-th/0309148}{{\tt hep-th/0309148}}].

\bibitem{Balthazar:2022apu}
B.~Balthazar, V.~A. Rodriguez, and X.~Yin, {\it {The S-Matrix of 2D Type 0B
  String Theory Part 2: D-Instanton Effects}},
  \href{http://arxiv.org/abs/2204.01747}{{\tt arXiv:2204.01747}}.

\bibitem{Chakravarty:2022cgj}
J.~Chakravarty and A.~Sen, {\it {Normalization of D instanton amplitudes in two
  dimensional type 0B string theory}},  {\em JHEP} {\bf 02} (2023) 170,
  [\href{http://arxiv.org/abs/2207.07138}{{\tt arXiv:2207.07138}}].

\bibitem{Sen:2022clw}
A.~Sen, {\it {Infrared finite semi-inclusive cross section in two dimensional
  type 0B string theory}},  \href{http://arxiv.org/abs/2208.07385}{{\tt
  arXiv:2208.07385}}.

\bibitem{Eniceicu:2022xvk}
D.~S. Eniceicu, R.~Mahajan, P.~Maity, C.~Murdia, and A.~Sen, {\it {The ZZ
  annulus one-point function in non-critical string theory: A string field
  theory analysis}},  {\em JHEP} {\bf 12} (2022) 151,
  [\href{http://arxiv.org/abs/2210.11473}{{\tt arXiv:2210.11473}}].

\bibitem{Sen:2004nf}
A.~Sen, {\it {Tachyon dynamics in open string theory}},  {\em Int. J. Mod.
  Phys. A} {\bf 20} (2005) 5513--5656,
  [\href{http://arxiv.org/abs/hep-th/0410103}{{\tt hep-th/0410103}}].

\bibitem{McGreevy:2003kb}
J.~McGreevy and H.~L. Verlinde, {\it {Strings from tachyons: The c=1 matrix
  reloaded}},  {\em JHEP} {\bf 12} (2003) 054,
  [\href{http://arxiv.org/abs/hep-th/0304224}{{\tt hep-th/0304224}}].

\bibitem{McGreevy:2003ep}
J.~McGreevy, J.~Teschner, and H.~L. Verlinde, {\it {Classical and quantum
  D-branes in 2-D string theory}},  {\em JHEP} {\bf 01} (2004) 039,
  [\href{http://arxiv.org/abs/hep-th/0305194}{{\tt hep-th/0305194}}].

\bibitem{Klebanov:2003km}
I.~R. Klebanov, J.~M. Maldacena, and N.~Seiberg, {\it {D-brane decay in
  two-dimensional string theory}},  {\em JHEP} {\bf 07} (2003) 045,
  [\href{http://arxiv.org/abs/hep-th/0305159}{{\tt hep-th/0305159}}].

\bibitem{Balthazar:2017mxh}
B.~Balthazar, V.~A. Rodriguez, and X.~Yin, {\it {The $c$ = 1 string theory
  S-matrix revisited}},  {\em JHEP} {\bf 04} (2019) 145,
  [\href{http://arxiv.org/abs/1705.07151}{{\tt arXiv:1705.07151}}].

\bibitem{Dorn:1994xn}
H.~Dorn and H.~J. Otto, {\it {Two and three point functions in Liouville
  theory}},  {\em Nucl. Phys.} {\bf B429} (1994) 375--388,
  [\href{http://arxiv.org/abs/hep-th/9403141}{{\tt hep-th/9403141}}].

\bibitem{Zamolodchikov:1995aa}
A.~B. Zamolodchikov and A.~B. Zamolodchikov, {\it {Structure constants and
  conformal bootstrap in Liouville field theory}},  {\em Nucl. Phys.} {\bf
  B477} (1996) 577--605, [\href{http://arxiv.org/abs/hep-th/9506136}{{\tt
  hep-th/9506136}}].

\bibitem{Balthazar:2018qdv}
B.~Balthazar, V.~A. Rodriguez, and X.~Yin, {\it {Long String Scattering in c
  $=$ 1 String Theory}},  {\em JHEP} {\bf 01} (2019) 173,
  [\href{http://arxiv.org/abs/1810.07233}{{\tt arXiv:1810.07233}}].

\bibitem{Ribault:2015sxa}
S.~Ribault and R.~Santachiara, {\it {Liouville theory with a central charge
  less than one}},  {\em JHEP} {\bf 08} (2015) 109,
  [\href{http://arxiv.org/abs/1503.02067}{{\tt arXiv:1503.02067}}].

\bibitem{Schomerus:2003vv}
V.~Schomerus, {\it {Rolling tachyons from Liouville theory}},  {\em JHEP} {\bf
  11} (2003) 043, [\href{http://arxiv.org/abs/hep-th/0306026}{{\tt
  hep-th/0306026}}].

\bibitem{Kostov:2005kk}
I.~K. Kostov and V.~B. Petkova, {\it {Bulk correlation functions in 2-D quantum
  gravity}},  {\em Theor. Math. Phys.} {\bf 146} (2006) 108--118,
  [\href{http://arxiv.org/abs/hep-th/0505078}{{\tt hep-th/0505078}}].

\bibitem{Zamolodchikov:2005fy}
A.~B. Zamolodchikov, {\it {Three-point function in the minimal Liouville
  gravity}},  \href{http://arxiv.org/abs/hep-th/0505063}{{\tt hep-th/0505063}}.
  [Theor. Math. Phys.142,183(2005)].

\bibitem{Harlow:2011ny}
D.~Harlow, J.~Maltz, and E.~Witten, {\it {Analytic Continuation of Liouville
  Theory}},  {\em JHEP} {\bf 12} (2011) 071,
  [\href{http://arxiv.org/abs/1108.4417}{{\tt arXiv:1108.4417}}].

\bibitem{McElgin:2007ak}
W.~McElgin, {\it {Notes on Liouville Theory at c \ensuremath{<}= 1}},  {\em
  Phys. Rev. D} {\bf 77} (2008) 066009,
  [\href{http://arxiv.org/abs/0706.0365}{{\tt arXiv:0706.0365}}].

\bibitem{Giribet:2011zx}
G.~Giribet, {\it {On the timelike Liouville three-point function}},  {\em Phys.
  Rev. D} {\bf 85} (2012) 086009, [\href{http://arxiv.org/abs/1110.6118}{{\tt
  arXiv:1110.6118}}].

\bibitem{Bautista:2019jau}
T.~Bautista, A.~Dabholkar, and H.~Erbin, {\it {Quantum Gravity from Timelike
  Liouville theory}},  {\em JHEP} {\bf 10} (2019) 284,
  [\href{http://arxiv.org/abs/1905.12689}{{\tt arXiv:1905.12689}}].

\bibitem{Balthazar:2022atu}
B.~Balthazar, V.~A. Rodriguez, and X.~Yin, {\it {The S-Matrix of 2D Type 0B
  String Theory Part 1: Perturbation Theory Revisited}},
  \href{http://arxiv.org/abs/2201.05621}{{\tt arXiv:2201.05621}}.

\bibitem{Zamolodchikov:1985ie}
A.~B. Zamolodchikov, {\it {CONFORMAL SYMMETRY IN TWO-DIMENSIONS: AN EXPLICIT
  RECURRENCE FORMULA FOR THE CONFORMAL PARTIAL WAVE AMPLITUDE}},  {\em Commun.
  Math. Phys.} {\bf 96} (1984) 419--422.

\bibitem{Collier:2023cyw}
S.~Collier, L.~Eberhardt, B.~Muehlmann, and V.~A. Rodriguez, {\it {The Virasoro
  minimal string}},  {\em SciPost Phys.} {\bf 16} (2024), no.~2 057,
  [\href{http://arxiv.org/abs/2309.10846}{{\tt arXiv:2309.10846}}].

\bibitem{CarneirodaCunha:2003mxy}
B.~Carneiro~da Cunha and E.~J. Martinec, {\it {Closed string tachyon
  condensation and world sheet inflation}},  {\em Phys. Rev. D} {\bf 68} (2003)
  063502, [\href{http://arxiv.org/abs/hep-th/0303087}{{\tt hep-th/0303087}}].

\bibitem{Strominger:2003fn}
A.~Strominger and T.~Takayanagi, {\it {Correlators in time - like bulk
  Liouville theory}},  {\em Adv. Theor. Math. Phys.} {\bf 7} (2003), no.~2
  369--379, [\href{http://arxiv.org/abs/hep-th/0303221}{{\tt hep-th/0303221}}].

\bibitem{Martinec:2003ka}
E.~J. Martinec, {\it {The Annular report on noncritical string theory}},
  \href{http://arxiv.org/abs/hep-th/0305148}{{\tt hep-th/0305148}}.

\bibitem{Hellerman:2007fc}
S.~Hellerman and I.~Swanson, {\it {Charting the landscape of supercritical
  string theory}},  {\em Phys. Rev. Lett.} {\bf 99} (2007) 171601,
  [\href{http://arxiv.org/abs/0705.0980}{{\tt arXiv:0705.0980}}].

\bibitem{Itzhaki:2021scf}
N.~Itzhaki, {\it {String Theory and The Arrow of Time}},  {\em JHEP} {\bf 03}
  (2021) 192, [\href{http://arxiv.org/abs/2101.10142}{{\tt arXiv:2101.10142}}].

\bibitem{Hashimoto:2022dro}
A.~Hashimoto, N.~Itzhaki, and U.~Peleg, {\it {A worldsheet description of
  instant folded strings}},  {\em JHEP} {\bf 02} (2023) 088,
  [\href{http://arxiv.org/abs/2209.04988}{{\tt arXiv:2209.04988}}].

\bibitem{Anninos:2021ene}
D.~Anninos, T.~Bautista, and B.~M\"uhlmann, {\it {The two-sphere partition
  function in two-dimensional quantum gravity}},  {\em JHEP} {\bf 09} (2021)
  116, [\href{http://arxiv.org/abs/2106.01665}{{\tt arXiv:2106.01665}}].

\bibitem{Suzuki:2021zbe}
K.~Suzuki and T.~Takayanagi, {\it {JT gravity limit of Liouville CFT and matrix
  model}},  {\em JHEP} {\bf 11} (2021) 137,
  [\href{http://arxiv.org/abs/2108.12096}{{\tt arXiv:2108.12096}}].

\bibitem{Kapec:2020xaj}
D.~Kapec and R.~Mahajan, {\it {Comments on the quantum field theory of the
  Coulomb gas formalism}},  {\em JHEP} {\bf 04} (2021) 136,
  [\href{http://arxiv.org/abs/2010.10428}{{\tt arXiv:2010.10428}}].

\bibitem{Alexandrov:2002fh}
S.~Y. Alexandrov, V.~A. Kazakov, and I.~K. Kostov, {\it {Time dependent
  backgrounds of 2-D string theory}},  {\em Nucl. Phys. B} {\bf 640} (2002)
  119--144, [\href{http://arxiv.org/abs/hep-th/0205079}{{\tt hep-th/0205079}}].

\bibitem{Alexandrov:2003uh}
S.~Alexandrov, {\it {Backgrounds of 2-D string theory from matrix model}},
  \href{http://arxiv.org/abs/hep-th/0303190}{{\tt hep-th/0303190}}.

\bibitem{Karczmarek:2003pv}
J.~L. Karczmarek and A.~Strominger, {\it {Matrix cosmology}},  {\em JHEP} {\bf
  04} (2004) 055, [\href{http://arxiv.org/abs/hep-th/0309138}{{\tt
  hep-th/0309138}}].

\bibitem{Craps:2005wd}
B.~Craps, S.~Sethi, and E.~P. Verlinde, {\it {A Matrix big bang}},  {\em JHEP}
  {\bf 10} (2005) 005, [\href{http://arxiv.org/abs/hep-th/0506180}{{\tt
  hep-th/0506180}}].

\bibitem{Craps:2006xq}
B.~Craps, A.~Rajaraman, and S.~Sethi, {\it {Effective dynamics of the matrix
  big bang}},  {\em Phys. Rev. D} {\bf 73} (2006) 106005,
  [\href{http://arxiv.org/abs/hep-th/0601062}{{\tt hep-th/0601062}}].

\bibitem{Banks:1996vh}
T.~Banks, W.~Fischler, S.~Shenker, and L.~Susskind, {\it {M theory as a matrix
  model: A Conjecture}},  {\em Phys. Rev. D} {\bf 55} (1997) 5112--5128,
  [\href{http://arxiv.org/abs/hep-th/9610043}{{\tt hep-th/9610043}}].

\bibitem{Polchinski:1998rq}
J.~Polchinski, {\em {String theory. Vol. 1: An introduction to the bosonic
  string}}.
\newblock Cambridge Monographs on Mathematical Physics. Cambridge University
  Press, 12, 2007.

\bibitem{Baumann:2022jpr}
D.~Baumann, D.~Green, A.~Joyce, E.~Pajer, G.~L. Pimentel, C.~Sleight, and
  M.~Taronna, {\it {Snowmass White Paper: The Cosmological Bootstrap}},  in
  {\em {2022 Snowmass Summer Study}}, 3, 2022.
\newblock \href{http://arxiv.org/abs/2203.08121}{{\tt arXiv:2203.08121}}.

\bibitem{Flauger:2022hie}
R.~Flauger, V.~Gorbenko, A.~Joyce, L.~McAllister, G.~Shiu, and E.~Silverstein,
  {\it {Snowmass White Paper: Cosmology at the Theory Frontier}},  in {\em
  {2022 Snowmass Summer Study}}, 3, 2022.
\newblock \href{http://arxiv.org/abs/2203.07629}{{\tt arXiv:2203.07629}}.

\bibitem{Moore:1988qv}
G.~W. Moore and N.~Seiberg, {\it {Classical and Quantum Conformal Field
  Theory}},  {\em Commun. Math. Phys.} {\bf 123} (1989) 177.

\bibitem{Yin:2017yyn}
X.~Yin, {\it {Aspects of Two-Dimensional Conformal Field Theories}},  {\em PoS}
  {\bf TASI2017} (2017) 003.

\bibitem{Hadasz:2009db}
L.~Hadasz, Z.~Jaskolski, and P.~Suchanek, {\it {Recursive representation of the
  torus 1-point conformal block}},  {\em JHEP} {\bf 01} (2010) 063,
  [\href{http://arxiv.org/abs/0911.2353}{{\tt arXiv:0911.2353}}].

\bibitem{Cho:2017oxl}
M.~Cho, S.~Collier, and X.~Yin, {\it {Recursive Representations of Arbitrary
  Virasoro Conformal Blocks}},  \href{http://arxiv.org/abs/1703.09805}{{\tt
  arXiv:1703.09805}}.

\bibitem{Chang:2014jta}
C.-M. Chang, Y.-H. Lin, S.-H. Shao, Y.~Wang, and X.~Yin, {\it {Little String
  Amplitudes (and the Unreasonable Effectiveness of 6D SYM)}},  {\em JHEP} {\bf
  12} (2014) 176, [\href{http://arxiv.org/abs/1407.7511}{{\tt
  arXiv:1407.7511}}].

\end{thebibliography}\endgroup

\end{document}